\documentclass[10pt]{article}
\usepackage[english]{babel}
\usepackage[utf8]{inputenc}
\usepackage[a4paper,top=2cm,bottom=2cm,left=3cm,right=3cm,marginparwidth=1.75cm]{geometry}

\usepackage[T1]{fontenc}
\usepackage{lmodern}

\usepackage{amsmath, amsfonts, amssymb}
\usepackage{mathtools}
\usepackage{graphicx}
\usepackage{algorithm}
\usepackage{algpseudocode}
\usepackage[colorlinks=true, urlcolor=black,linkcolor=blue, citecolor=blue]{hyperref}

\usepackage[auth-sc]{authblk}
\usepackage{orcidlink}

\usepackage{caption}
\usepackage{subcaption}

\usepackage{afterpage}
\usepackage{sectsty}
\usepackage[xindy,acronym]{glossaries}

\DeclareMathOperator*{\argmin}{arg\,min}
\makenoidxglossaries
\newacronym{tof}{ToF}{time-of-flight}
\newacronym{rdp}{RDP}{relative difference prior}
\newacronym{pet}{PET}{positron emission tomography}
\newacronym{petric}{PETRIC}{PET Rapid Image Reconstruction Challenge}
\newacronym{mlem}{MLEM}{maximum likelihood expectation maximization}
\newacronym{osem}{OSEM}{ordered subsets expectation maximization}
\newacronym{sirf}{SIRF}{Synergistic Image Reconstruction Framework}
\newacronym{sgd}{SGD}{Stochastic Gradient Descent}
\newacronym{saga}{SAGA}{Stochastic Averaged Gradient Amelior\'e}
\newacronym{svrg}{SVRG}{Stochastic Variance Reduced Gradient}
\newacronym{map}{MAP}{maximum a-posteriori}
\newacronym{bb}{BB}{Barzilai--Borwein}
\newacronym{nrmse}{NRMSE}{normalized root mean square error}
\newacronym{kl}{KL}{Kullback--Leibler}

\newcommand{\formula}[1]{#1}
\newcommand{\image}{\formula{x}}
\newcommand{\forwardop}{\formula{A}}
\newcommand{\iter}[1]{\formula{\image^{(#1)}}}
\newcommand{\precond}{\formula{D}}
\newcommand{\preconditer}[1]{\formula{D^{(#1)}}}
\newcommand{\data}{\formula{y}}
\newcommand{\datafit}{\formula{\mathcal D}}
\newcommand{\background}{\formula{r}}
\newcommand{\prior}{\formula{\mathcal R}}
\newcommand{\smoothness}{\formula{\mathcal S}}
\newcommand{\nonneg}{\formula{\iota_{\geq0}}}
\newcommand{\nonnegprox}{\formula{\operatorname{prox}_{\nonneg}}}
\newcommand{\diag}{\formula{\operatorname{diag}}}
\newcommand{\stepsize}[1]{\formula{\tau^{(#1)}}}
\newcommand{\objective}{\formula{\mathcal J}}
\newcommand{\nsubsets}{\formula{n}}
\newcommand{\gradientestimator}[1]{\formula{\tilde \nabla^{(#1)}}}
\newcommand{\regparam}{\formula{\tilde \beta}}
\newcommand{\regparamfinal}{\formula{\beta}}
\newcommand{\sagahist}[2]{\formula{g^{(#1)}_{#2}}}
\newcommand{\svrgref}{\formula{\hat \image}}
\newcommand{\svrghist}[1]{\formula{\hat g_{#1}}}

\newcommand{\N}{\formula{\mathbb{N}}}

\allsectionsfont{\sffamily}
\usepackage{float}
\floatstyle{boxed}

\restylefloat*{figure}
\restylefloat*{table}

\usepackage{enumitem}

\setlist[itemize]{
  topsep=2pt,      
  parsep=0.5pt,   
}

\setlist[itemize,1]{label=\scriptsize\textbullet}

\setlist[itemize,2]{label=\tiny\textbullet}

\usepackage{booktabs}
\usepackage{makecell}

\makeatletter
\patchcmd{\@maketitle}
  {\@date}                        % original
  {{\sffamily\@date}}             % replacement
  {}{}                            % success/failure hooks
\makeatother

\makeatletter
\renewenvironment{abstract}
  {\par\noindent\textbf{\textsf{Abstract:}}\ \ignorespaces}
  {\par\medskip}
\makeatother

\title{\Large\sffamily\textbf{Fast PET Reconstruction with Variance Reduction and Prior-Aware Preconditioning}}

\author[1]{Matthias~J.~Ehrhardt \orcidlink{0000-0001-8523-353X}}
\author[2]{Zeljko~Kereta \orcidlink{0000-0003-2805-0037}}
\author[3]{Georg~Schramm \orcidlink{0000-0002-2251-3195}}

\affil[1]{Department of Mathematical Sciences, University of Bath, UK}
\affil[2]{Computer Science Department, University College London, UK}
\affil[3]{Department of Imaging and Pathology, KU Leuven, Belgium}

\begin{document}
\maketitle

\begin{abstract}
We investigate subset-based optimization methods for \gls{pet} image reconstruction incorporating a regularizing prior. 
\gls{pet} reconstruction methods that use a prior, such as the \gls{rdp}, are of particular relevance, as they are widely used in clinical practice and have been shown to outperform conventional early-stopped and post-smoothed \gls{osem}.

Our study evaluates these methods on both simulated data and real brain \gls{pet} scans from the 2024 \gls{petric}, where the main objective was to achieve \gls{rdp}-regularized reconstructions as fast as possible, making it an ideal benchmark. Our key finding is that incorporating the effect of the prior into the preconditioner is crucial for ensuring fast and stable convergence.

In extensive simulation experiments, we compare several stochastic algorithms---including \gls{sgd}, \gls{saga}, and \gls{svrg}---under various algorithmic design choices and evaluate their performance for varying count levels and regularization strengths. 
The results show that \gls{svrg} and \gls{saga} outperformed \gls{sgd}, with \gls{svrg} demonstrating a slight overall advantage.
The insights gained from these simulations directly contributed to the design of our submitted algorithms, which formed the basis of the winning contribution to the \gls{petric} 2024 challenge.
\end{abstract}

\section{Introduction}

\subsection{Context}
\Gls{pet} is a pillar in modern clinical imaging widely used in oncology, neurology and cardiology. 
Most state-of-the-art approaches for the image reconstruction problem in \gls{pet} imaging can be cast as an optimization problem
\begin{align}\label{eqn:optimisation_problem}
    \image^* \in \argmin_{\image} \left\{ \datafit(\forwardop\image + \background, \data) + \prior(\image)\right\},
\end{align}
where the data-fidelity term $\datafit : \mathcal Y \times \mathcal Y \to [0, \infty]$ measures how well the estimated data $\forwardop\image + \background$ matches the acquired data $\data$ and the regularizer $\prior : \mathcal X \to [0, \infty]$ penalizes unwanted features in the image. $\forwardop : \mathcal X \to \mathcal Y$ is a linear forward model for the \gls{pet} physics, including effects such as scanner sensitivities or attenuation, and $\background$ is the additive background term to account for scattered and
random coincidences. Due to the Poisson nature of the data, the data-fidelity is usually taken as the \gls{kl} divergence. The regularizer commonly entails nonnegativity constrains and terms promoting smoothness. A particularly successful model for smoothness in \gls{pet} is the \gls{rdp}~\cite{nuyts2002concave}. 

This paper is concerned with algorithms for a fast reconstruction of $\image^*$. 
Particularly, we present our winning contribution to the 2024 image reconstruction challenge \gls{petric} \cite{petric}, where the task was to reconstruct data acquired by a range of \gls{pet} scanners using \gls{rdp} regularized reconstruction methods. 
\gls{pet} image reconstructions that use the \gls{rdp} are of particular current relevance, as \gls{rdp} is widely used in clinical practice, being implemented by a major commercial vendor, and has been shown to outperform conventional early-stopped and post-smoothed OS-MLEM reconstructions~\cite{teoh2015, teoh2015b, Ahn2015}.
Their implementation is based on BSREM \cite{ahn2003globally}. It was shown to be outperformed in terms of speed by an algorithm using ideas from machine learning and a tailored preconditioning \cite{twyman2022investigation}. 
In this paper, we outline our process in finding the winning algorithm and share the insights we gained along the way. 
For context, the task had to be completed in \gls{sirf} \cite{ovtchinnikov2020sirf} and speed was measured as walltime until an application-focused convergence criteria were reached. 

\subsection{Problem Details}
Fast algorithms for \gls{pet} reconstruction have traditionally been subset-based \cite{hudson1994accelerated}, that is, only a subset of the data is used in every iteration. 
Over the last decade, algorithms using a similar strategy but derived for machine learning have entered the field, showing state-of-the-art performance~\cite{twyman2022investigation, ehrhardt2019faster, kereta2021stochastic, schramm2022fast, ehrhardt2024guide, papoutsellis2024stochastic}. 
They exploit the fact that the \gls{kl} divergence is separable in the estimated data
\begin{align}\label{eqn:separable_datafit}
    \datafit(\forwardop\image + \background, \data) = \sum_{i=1}^\nsubsets \sum_{j \in S_i} 
 d(\forwardop_j \image + \background_j, \data_j),
\end{align}
where $\nsubsets$ denotes the number of subsets and  function $d$ is defined by
\begin{align*}
    d(s, t) = \begin{cases} 
    s - t + t \log(t/s), & \text{if $t > 0, s > 0$} \\
    s, & \text{if $t = 0, s \geq 0$} \\
    \infty, &\text{otherwise}
    \end{cases}.
\end{align*}
Here $S_i$ denote a subset of the data, e.g., all data associated to a ``view''. 

A lot of effort has been put into finding good prior models (i.e., regularizers) for PET, including smooth and nonsmooth priors, promoting smoothness of the image to be reconstructed or promoting similarity to anatomical information \cite{bredies2010total, knoll2016joint, ehrhardt2016pet, xie2020generative, singh2024scorebasedpet}. In \cite{nuyts2002concave}, the authors propose a smooth and convex prior that takes into account the scale of typical \gls{pet} images, resulting in promoting more smoothness in less active regions. Mathematically for nonnegative images $\image$ the resulting regularizer can be defined by
\begin{align}\label{eqn:rdp}
    \smoothness(\image) = \frac{1}{2} \sum_i \sum_{j \in N_i} w_{i,j} \kappa_i \kappa_j \frac{(\image_i - \image_j)^2}{\image_i + \image_j + \gamma |\image_i - \image_j| + \varepsilon},
\end{align}
where the first sum is over all voxels $i$ and the second sum is over all ``neighbors'' $j$. The parameter $\gamma > 0$ allows placing more or less emphasis on edge-preservation and the parameter $\varepsilon > 0$ ensures that the function is well-defined and twice continuously differentiable. The terms $w_{i,j}$, $\kappa_i$ and $\kappa_j$ are weight factors accounting for distances between voxels and are intended to create a uniform ``perturbation response'' \cite{tsai2019benefits}. Note that the essential part of the prior is
\begin{align*}
    \phi(s,d) = \frac{d^2}{s + \gamma |d| + \varepsilon},
\end{align*}
which has two important properties. First, if the sum of activities between voxels $s$ is small compared to the scaled absolute difference $\gamma |d|$, the regularizer essentially reduces to total variation: $\phi(s,d) \approx |d| / \gamma$. Second, the larger the activity in both voxels, i.e., the larger $s$, the less weight is given on penalizing their difference, justifying the name of the regularizer. See also Appendix \ref{appendix:rdp} for formulas of derivatives.

Combined with the indicator function of the nonnegativity constraint, 
$$
\nonneg(\image) = \begin{cases} 0, & \text{if $\image_i \geq 0$ for all $i$} \\
\infty, & \text{otherwise}
\end{cases},
$$
we arrive at the regularization model used in \gls{petric}
\begin{align} \label{eqn:prior}
    \prior(\image) = \regparamfinal \smoothness(\image) + \nonneg(\image).
\end{align}
This formula has to be interpreted to be $\infty$ for infeasible images with negative voxel-values and has the finite \gls{rdp} value everywhere else.

The rest of the paper is structured as follows. In section \ref{sec:building_blocks} we introduce the building blocks of our algorithms. We discuss proximal stochastic gradient approaches for the solution of \eqref{eqn:optimisation_problem}, the stepsize regimes, preconditioning and subset selection. In section \ref{sec:simulation_experiments} we thoroughly investigate the effects of different choices of building blocks in a simulated setting. In section \ref{sec:submitted_algorithms} we present the algorithms we ended up using in PETRIC and their performance on real data. We conclude in sections \ref{sec:discussion} and \ref{sec:conclusions} with final remarks.

\section{Building Blocks}\label{sec:building_blocks}

Combining the modeling choices in \eqref{eqn:optimisation_problem}, \eqref{eqn:separable_datafit} and \eqref{eqn:prior}, we arrive at the optimization problem
\begin{align} \label{eqn:template}
    \min_{\image} \left\{ \sum_{i=1}^\nsubsets \objective_i(\image) + \nonneg(\image)\right\},
\end{align}
where we define $\objective_i(\image) = \datafit_i(\image) +\frac{\regparamfinal}{\nsubsets} \smoothness(\image)$, and $\datafit_i(\image)\coloneq \sum_{j \in S_i} d(\forwardop_j \image + \background_j, \data_j)$.
The zoo of optimization methods for solving instances of problem \eqref{eqn:template} is rich and has been growing in recent decades, see~\cite{ehrhardt2024guide} and references therein. 
For linear inverse problems, such as in \gls{pet} image reconstruction, the most common approaches are based on (proximal) gradient descent or on primal-dual approaches.

In this work we consider stochastic gradient methods for the solution of the problem \eqref{eqn:template}.
They take the form
\begin{equation}\label{eqn:sgm}
    \iter{k+1} = \nonnegprox\big(\iter{k} - \stepsize{k} \preconditer{k}\gradientestimator{k} \big),
\end{equation}
where $\stepsize{k} >0$ is a stepsize, $\gradientestimator{k}$ is an estimator of the gradient of the smooth part of the objective function $\objective(\image)= \sum_{i=1}^\nsubsets \objective_i(\image)$, $\preconditer{k}$ is a matrix that acts as a preconditioner, and $\nonnegprox$ is the proximal operator associated with the nonnegativity constraint which can be efficiently computed entry-wise, $[\nonnegprox(\image)]_j = \max(0,\image_j)$.

All three components $\gradientestimator{k}$, $\preconditer{k}$ and $\stepsize{k}$ are critical for fast and stable algorithmic performance.
In realistic image reconstruction settings, and in the context of the \gls{petric} challenge, the selection of these three components must balance accuracy and computational costs.
In the remainder of this section, we review stochastic estimators and discuss their tradeoffs, address the stepsize selection and preconditioners.
Lastly, we consider the role of subset selection and sampling regimes. Namely, how to choose the sets $S_i$ in \eqref{eqn:separable_datafit} and decide which subsets to use at each iteration of the algorithm.

\subsection{Stochastic Gradient Methods}

Let's turn our attention to the selection of gradient estimators $\gradientestimator{k}$.

\textbf{\acrlong{sgd} (\acrshort{sgd})}
defines the gradient estimator by selecting a random subset index $i_k$ in each iteration and evaluating 
\begin{equation*}
\gradientestimator{k} := \nsubsets\nabla \objective_{i_k}(\iter{k})
\end{equation*}
to compute the update \eqref{eqn:sgm}.
Each iteration requires storing only the current iterate and computing the gradient of only one subset function.
This can lead to large variances across updates, which increase with the number of subsets. To moderate this,  vanishing stepsizes, satisfying $$\sum_{k=1}^\infty \stepsize{k} = \infty \text{ and }\sum_{k=1}^\infty (\stepsize{k})^2 < \infty,$$ are required to ensure convergence, but at the cost of convergence speed.

\textbf{\acrlong{saga} (\acrshort{saga})} \cite{Defazio2014}
controls the variance by keeping a table of historical gradients $(\sagahist{k}{i})_{i=1}^\nsubsets \in \mathcal X^\nsubsets$. 
Each iteration uses a computed subset gradient combined with the full gradient table to update the gradient estimator
\begin{align}\label{eqn:saga}
    \gradientestimator{k} = \nsubsets(\nabla \objective_{i_k}(\iter{k}) - \sagahist{k}{i_k})+\sum_{i=1}^{\nsubsets} \sagahist{k}{i},
\end{align}
followed by updating the corresponding entry in the table
$$
\sagahist{k+1}{j} = \begin{cases}
    \nabla \objective_{i_k}(\iter{k}), &\text{if $j = i_k$} \\
    \sagahist{k}{j}, &\text{otherwise} 
\end{cases}.
$$
In contrast to \gls{sgd}, \gls{saga} guarantees convergence to a minimizer with constant stepsizes and preconditioners for Lipschitz-smooth problems.
In its standard form \gls{saga} has the same computational cost as \gls{sgd}, but requires storing $\nsubsets$ gradients. 
The memory cost is not a practical limitation for most \gls{pet} problems (even for relatively large $\nsubsets$). If this is a concern, alternative formulations of \gls{saga} exist with other memory footprints, see \cite{ehrhardt2024guide} for a further discussion.

\textbf{\acrlong{svrg} (\acrshort{svrg})} \cite{johnson2013svrg} reduces the variance by storing reference images and gradients. In contrast to \gls{saga}, these are updated infrequently. \gls{svrg} is usually implemented with two loops: an outer loop and an inner loop.
At the start of each outer loop subset gradients and the full gradient estimator are computed at the last iterate as $$\svrghist{i} = \nabla \objective_i(\svrgref), \quad \svrghist{} = \sum_{i=1}^\nsubsets\svrghist{i}.$$ 
In the inner loop the gradients are retrieved from memory and balanced against a randomly sampled subset gradient at the current iterate, giving the gradient estimator \begin{align}\label{eqn:svrg}
    \gradientestimator{k} = \nsubsets(\nabla \objective_{i_k}(\iter{k}) - \svrghist{i_k}) +\svrghist{}. 
\end{align}
Note the similarity between the gradient estimators of \gls{saga} and \gls{svrg} given by \eqref{eqn:saga} and \eqref{eqn:svrg}, respectively.
After $\omega \nsubsets$ iterations the snapshot image and the full gradient estimator are updated.
The update parameter $\omega\in\N$ is typically chosen as $2$ for convex problems.

It is most common to store only the snapshot image $\svrgref$ and the corresponding full gradient $\sum_{i=1}^{\nsubsets} \svrghist{i}$, which then requires recomputing the subset gradient $\svrghist{i_k}$ at each iteration. This lowers the memory footprint (requiring only the snapshot image and the full gradient to be stored), but increases the computational costs.

\subsection{Stepsizes}

Theoretical convergence guarantees often require stepsizes based on $L_{\rm{max}} = {\rm{max}}_{i=1,\ldots,\nsubsets} \{L_i\}$, where $L_i$ is the Lipschitz constant of $\nabla \objective_i$. 
In \gls{pet},  global Lipschitz constants are usually pessimistic, yielding conservative stepsize estimates. 

Many stepsize approaches exist for stochastic iterative methods, ranging from predetermined choices made before running the algorithm (constant or vanishing), to adaptive methods (e.g., \gls{bb} \cite{tan2016barzilai} and ``difference of gradients''-type \cite{ivgi2023dog} rules), and backtracking techniques (e.g., Armijo \cite{vaswani2019painless}).
Due to the constraints imposed by the challenge (where computational time is a key metric), in this work we focus on the first two categories.

\textbf{Constant} is the baseline stepsize rule. The specific value requires tuning to ensure convergence.
    
\textbf{Vanishing} rules consider stepsizes of the form $\stepsize{k}= \stepsize{0} / (1+\eta k/\nsubsets)$, which satisfy \gls{sgd} convergence conditions, for $\stepsize{0} >0$ and decay parameter $\eta>0$ that needs balance convergence and stability: small enough to maintain speed but large enough to ensure convergence.
    
\textbf{Adaptive} stepsize tuning via the \gls{bb} rule is achieved by minimizing the residual of the secant equation at the current iterate. 
It converges for strongly convex problems and it is applicable to \gls{sgd} and \gls{svrg} \cite{tan2016barzilai}. 
We experimented with several variants (long and short forms, geometric mean combinations, diagonal \gls{bb}, etc.) but settled on the short form \gls{bb} for performance and stability. 
When applied to gradient descent, short form BB sets the stepsizes according to $\stepsize{k} = p^\top q/ (q^\top q)$, where $p =\iter{k}-\iter{k-1}$ and $q=\gradientestimator{k} -\gradientestimator{k-1}$.  
When applied to SVRG these values are computed in iterations when the full gradient is recomputed.

\subsection{Preconditioning}
Preconditioners are essential for accelerating iterative reconstruction algorithms by stabilizing admissible stepsize and adapting them to individual components of the solution. 
Effectively, image components with large gradient variance get smaller updates, and vice versa. 
This can have a dramatic effect in \gls{pet} image reconstruction (and machine learning applications) due to wildly varying range of local Lipschitz constants. 
Motivated by Newton's method, many preconditioners aim to approximate the inverse of the Hessian and thus may allow unit stepsizes. However, computing full Hessians is impractical in high dimensions, motivating the need for efficient approximations.

Preconditioners based on only the data-fidelity are standard in \gls{pet}. 
The most prominent example is $$\precond_{\rm MLEM}(\image)= \diag\left(\frac{\image + \delta}{\forwardop^\top 1}\right),$$ which can be derived from the gradient descent interpretation of \gls{mlem}. 
Here, the division of the two vectors is interpreted componentwise. 
Since $\image \geq 0$ and $\forwardop^\top 1 > 0$, a small constant $\delta > 0$ ensures that the every diagonal entry of the preconditioner is non-zero. 
$\precond_{\rm MLEM}$ tends to work well for weak priors (e.g., in low-noise scenarios).
However, it often underperforms as it does not account for the strength of the prior.
This can either jeopardize the convergence behavior or require significant stepsize tuning. 

Let $$\precond_{\regparamfinal\smoothness}(\image)= \diag\left(\frac{1}{\diag (H_{\regparamfinal\smoothness}(\image))}\right)$$ be the inverse of the diagonal of the Hessian of the regularizer.
In this work we use diagonal preconditioners that combine the data-fidelity and the prior terms via the (scaled) harmonic mean between $\precond_{\rm MLEM}$ and $\precond_{\regparamfinal\smoothness}$. 
For scalars $a, b > 0$, the harmonic mean is given by
$$
h(a, b) = \frac{2}{\frac{1}{a} + \frac{1}{b}}.
$$
Since our preconditioners are diagonal, this can be readily extended to define for some $\alpha > 0$
\begin{align}\label{eqn:preconditioner}
    \precond(\image) 
    &= \frac12 h(\precond_{\rm MLEM}(\image), \alpha^{-1} \precond_{\regparamfinal\smoothness}(\image)) \notag \\
    &= \left(\precond_{\rm MLEM}^{-1}(\image)+\alpha \precond^{-1}_{\regparamfinal\smoothness}(\image)\right)^{-1} \notag \\
    &= \diag \left(\frac{\image+\delta}{\forwardop^\top1+\alpha{\diag(H_{\regparamfinal\smoothness}(\image))(\image+\delta)}}\right).
\end{align}
Note that it satisfies $\precond(\image)\leq \min\{\precond_{\rm MLEM}(\image), \alpha^{-1} \precond_{\regparamfinal\smoothness}(\image)\}$.
While this may look like an ad-hoc choice, if $\precond_{\rm MLEM}$ and $\alpha^{-1}\precond_{\regparamfinal\smoothness}$ are good approximations to their respective Hessians, then the harmonic mean $\precond$ will be a good approximation to Hessian of the entire smooth term $\objective$. 

We tested several alternatives to \eqref{eqn:preconditioner}, such as taking an componentwise minimum between $\precond_{\rm MLEM}$ and $\precond_{\regparamfinal\smoothness}$, reweighing their contributions, using the Kailath variant of the Woodbury identity (together with the diagonal approximation) to estimate the inverse of the Hessian, and other variants.
The selected preconditioner provided the best compromise between computational costs required to compute it and algorithmic performance.
Traditional second order methods update the preconditioner in every iteration, which is costly.
Preconditioner \eqref{eqn:preconditioner} is much cheaper and, as experiments show,  requires updating only in the first 3 epochs, after which it stabilizes with no performance gain from further updates.

\subsection{Subset Selection and Sampling}
Subset-based reconstruction algorithms enhance the convergence speed of traditional iterative methods by dividing the projection data into multiple subsets and performing updates using partial measurement data. 
While this approach can offer significant computational advantages, careful selection of the number of subsets is critical. 
Using too many subsets can introduce artifacts and amplify noise, especially when subsets lack sufficient angular coverage, and increases the variance between successive updates, which can compromise the stability and convergence properties.
Conversely, selecting too few subsets diminishes the acceleration benefit and causes behavior similar to classical methods, such as \gls{mlem}, which are known for their slow convergence. 
The number of subsets $\nsubsets$ is typically chosen as a divisor of the total number of projection angles (or views), allowing the data to be partitioned evenly. Subsets are then constructed to ensure that each is representative and uniformly distributed.
We found that using approximately $25$ subsets provides a good tradeoff between reconstruction quality and computational speed in most scenarios, given the current computational requirements and scanner configurations.

To determine the order in which subsets are accessed we consider the following standard choices

\textbf{Herman--Meyer order \cite{herman1993hmorder}} is a well-established deterministic choice that is based on the prime decomposition of the number of subsets.

\textbf{Uniformly random with replacement} is the most common choice in machine learning applications. In each iteration the subset index $i$ is chosen by taking a sample from $\{1,\ldots, \nsubsets\}$ uniformly at random.

\textbf{Uniformly random without replacement} randomizes access to subset indices but ensures $\nsubsets$ successive iterates cycle through all the data by computing a permutation of $(1,\ldots,\nsubsets)$ in each epoch.

\textbf{Importance sampling} uses a weighted variant of uniform sampling with replacement. For each $1 \leq i \leq \nsubsets$ we assign a probability $p_i\geq0$, such that $\sum_{i=1}^\nsubsets p_i=1$. When Lipschitz constants $L_i$ are known then $p_i = L_i / \sum_{j=1}^\nsubsets L_j$ is a common choice.

Since Lipschitz constants $L_i$ are unknown in \gls{pet}, we propose an alternative importance sampling strategy for \gls{svrg}. Namely, when the full gradient estimator is updated we compute $p_i = \|\nabla \objective_i(\image)\| / \sum_{j=1}^\nsubsets \|\nabla \objective_j(\image)\|$, where $\image$ is the current image estimate. This incurs minimal computational overhead, since in \gls{svrg} all the subset gradients are already recomputed. 

Lastly, drawing inspiration from the Herman--Meyer ordering, which is designed to maximize information gain between successive updates, and incorporating the concept of random sampling without replacement to ensure full coverage of subsets in each epoch with varying order, we propose the following novel subset ordering strategy.

\textbf{Cofactor order} begins by identifying all generators of the cyclic group associated with the number of subsets, $\nsubsets$, which are identified as positive integers $k<\nsubsets$ that are coprime with $\nsubsets$, meaning they share no common prime factors with it.
These generators are then ranked according to their proximity to two reference points: $0.3 \nsubsets$ and $0.7 \nsubsets$, aiming to balance spread and randomness. In each epoch, the next available generator from this sorted list is selected and used to define a new traversal of the cyclic group, thereby determining the order in which subsets are accessed (i.e., one subset index per iteration). Once the list of generators is exhausted, it is reinitialized, and the process repeats for subsequent epochs.

\begin{figure}
    \centering
    \includegraphics[width=0.95\linewidth]{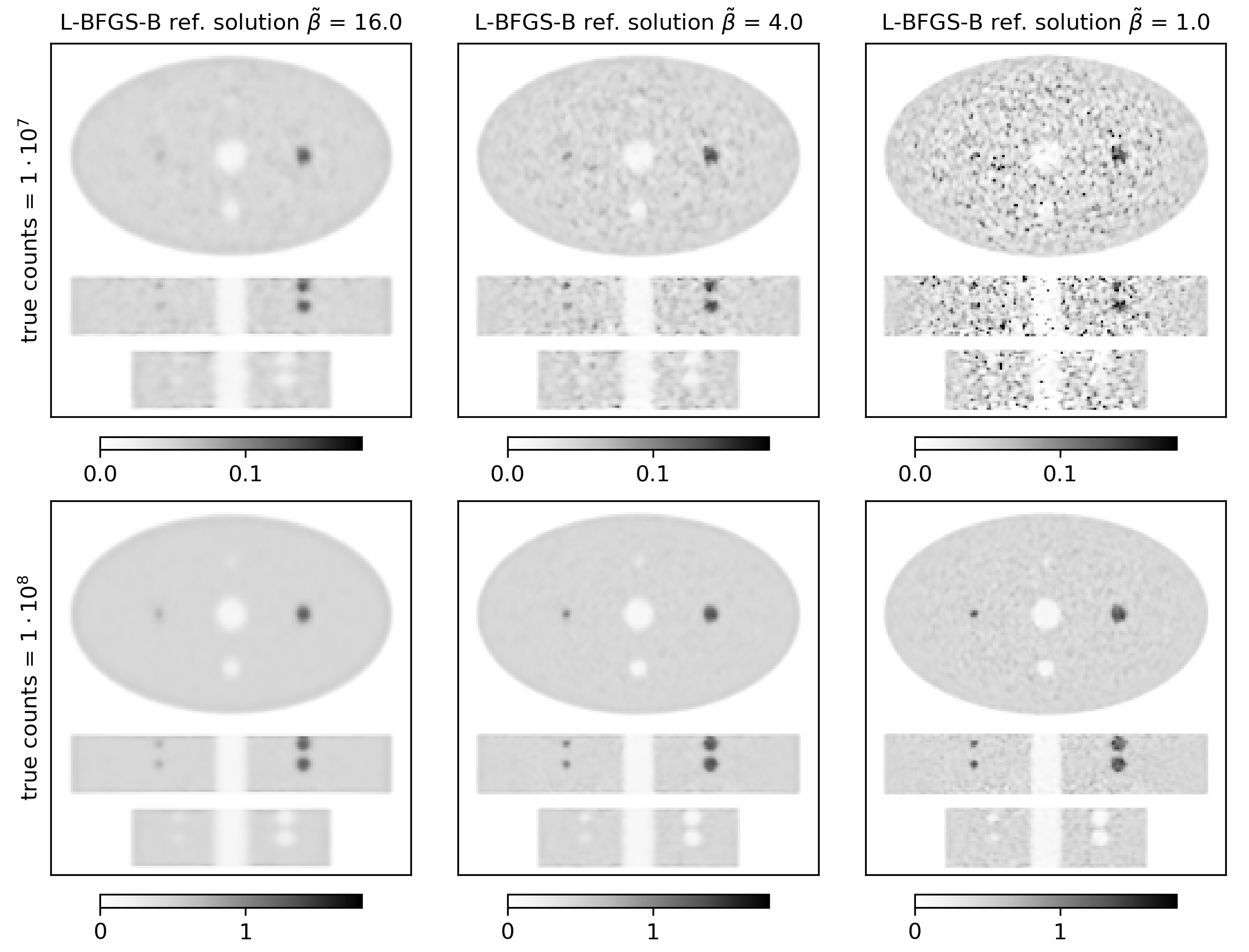}
    \caption{Stacked central transversal, coronal and sagittal slices of L-BFGS-B reference reconstructions of the ellipse phantom.
     Each column shows a different level of regularization ($\tilde{\beta}$) increasing from left to right.
     Top row shows results for $10^7$ true counts, and bottom row for $10^8$ true counts.}
    \label{fig:ref-recons}
\end{figure}

\section{Numerical Simulation Experiments}\label{sec:simulation_experiments}

To validate and refine the algorithmic components introduced in the previous section, we conducted a comprehensive suite of fast \emph{inverse‐crime} simulations. 
By simulating a simplified yet realistic \gls{pet} scanner using the pure GPU mode of \texttt{parallelproj v1.10.1} \cite{schramm2023}, iterative reconstructions could be run in seconds.
This enabled a systematic exploration of the effects of various factors on convergence behavior, including the choice of stochastic algorithm, preconditioner, step‐size strategy, number of subsets, subset sampling method, \gls{tof} versus non‐\gls{tof} data, count levels, and regularization strength.

\subsection{Simulation Setup}
All experiments used a simulated cylindrical (polygonal) scanner with a diameter of 600\,mm and a length of 80\,mm, composed of 17 rings with 36 modules each (12 detectors per module). 
Simulated \gls{tof} resolution was 390\,ps, and a 4\,mm isotropic Gaussian kernel in image space was used to model limited spatial resolution. 
Emission data was binned into a span 1 sinogram  (289 planes, 216 views, 353 radial bins, 25 \gls{tof} bins). 
A simple 3D elliptical phantom was forward‐projected (accounting for water attenuation), contaminated with a smooth background sinogram, and corrupted by Poisson noise to simulate realistic emission data. 
Low and high count regimes were simulated with $10^7$ and $10^8$ true events, respectively.
Reconstruction was performed at image size $161\times161\times33$ voxels with a 2.5\,mm isotropic spacing.

Reference reconstructions (see Fig.~\ref{fig:ref-recons}) were obtained by running 500 iterations of preconditioned L-BFGS-B with three relative regularization strengths $\regparam\in\{1,4,16\}$. 
The regularization parameter $\regparamfinal$ was scaled as
\begin{equation}
  \regparamfinal = \regparam \times 2\times10^{-4} \times \frac{\text{true counts}}{3\times10^7}\,.
\end{equation}
This ensures that reconstructions with the same $\regparam$ at different count levels show comparable resolution. 
All stochastic reconstructions were initialized with one epoch of \gls{osem} (with 27 subsets). 
Convergence was measured by the \gls{nrmse} excluding cold background around the elliptical phantom, normalized by the intensity of the largest background ellipsoid. 
In line with the \gls{nrmse} target threshold used in the \gls{petric} challenge, we consider the point where \gls{nrmse} was less than 0.01 as a marker of practical convergence.
The data was divided into $\nsubsets$ subsets by selecting every $\nsubsets$-th view. Unless stated otherwise, in each epoch subsets were drawn uniformly at random without replacement.
All runs were performed using an NVIDIA RTX A4500 GPU. The code for all our simulation experiments as well as our submissions to \gls{petric} is available on \href{https://github.com/SyneRBI/PETRIC-MaGeZ/tree/e355220aebbd8f3ce66315a8bd4b5ddc324fa578}{GitHub}.

\begin{figure}
    \centering
    \subfloat[\gls{svrg}]{%
    \includegraphics[width=0.99\linewidth]{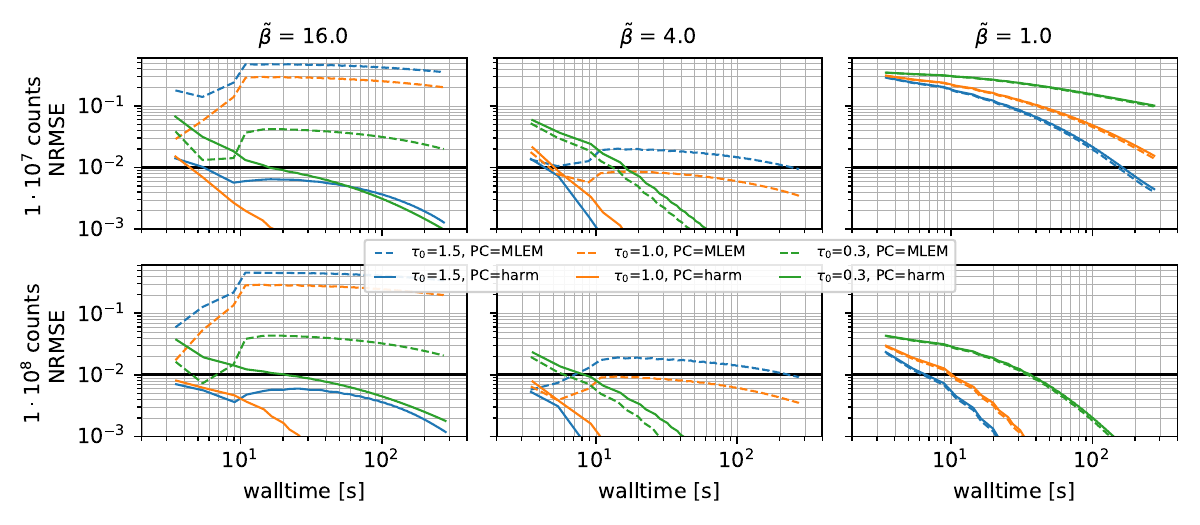}%
        \label{fig:pc_SVRG}
    }\\
    \subfloat[\gls{saga}]{%
        \includegraphics[width=0.99\linewidth]{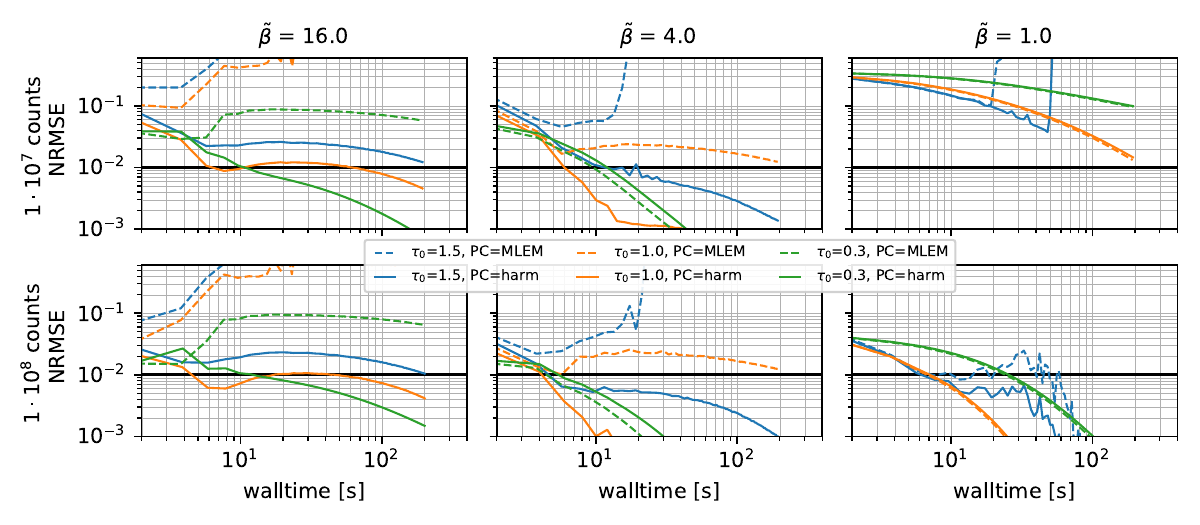}%
        \label{fig:pc_SAGA}
    }\\
    \subfloat[\gls{sgd}]{%
        \includegraphics[width=0.99\linewidth]{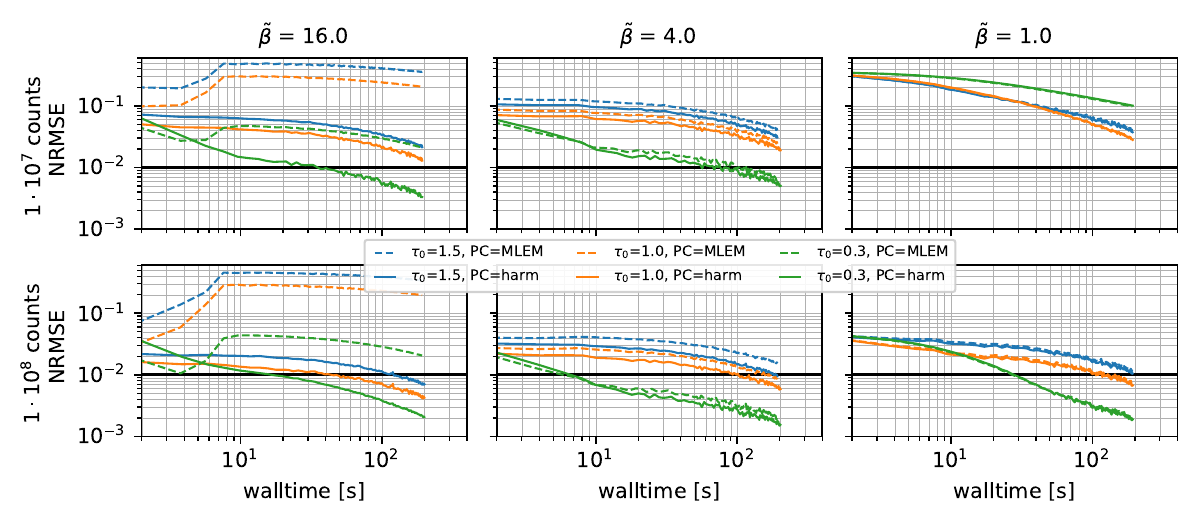}%
        \label{fig:pc_SGD}
    }
    \caption{Reconstruction performance in terms of \gls{nrmse} versus walltime for \textbf{\gls{svrg}, \gls{saga}, \gls{sgd}}, for \gls{mlem} (dashed lines) and harmonic (solid lines) \textbf{preconditioners} (PC) and three \textbf{initial stepsizes} ($\stepsize{0}$) represented by different colors, using 27 subsets, a gentle stepsize decay with $\eta = 0.02$, 100 epochs, and subset selection without replacement.
    Results are shown for three levels of regularization ($\regparam$) and two count levels. 
    \textbf{Note the logarithmic scale on the x and y axes.}
    For each combination of $\nsubsets$ and $\stepsize{0}$, the outcome of \textbf{1 run} is displayed.
    The thick horizontal black line shows the \gls{nrmse} target threshold of $10^{-2}$ used in \gls{petric}.}
    \label{fig:pc}
\end{figure}

\subsection{Main Simulation Results}

\paragraph{Algorithm and preconditioner effects (see Fig.~\ref{fig:pc}):}
When comparing \gls{svrg}, \gls{saga} and plain \gls{sgd} under a vanishing stepsize schedule $\stepsize{k}=\stepsize{0}/(1+0.02\,k/\nsubsets)$ with $\stepsize{0}\in\{0.3,1.0,1.5\}$ and $\nsubsets=27$, we made the following observations.
\begin{itemize}
  \item \gls{svrg} and \gls{saga} consistently outperform \gls{sgd} in all count and regularization regimes.
  \item The harmonic‐mean preconditioner \eqref{eqn:preconditioner} is crucial: under strong regularization $\regparam = 16$, the classic \gls{mlem} preconditioner diverges or converges extremely slowly (depending on the chosen stepsize), whereas the harmonic‐mean variant converges reliably in every scenario.
  \item \gls{svrg} with the harmonic preconditioner, $\stepsize{0}=1$ and $\eta=0.02$ (giving mild decay) yields the fastest convergence for medium and high $\regparam$. 
  For low regularization, a slightly larger $\stepsize{0}$ (up to 1.5 or 2.5) can accelerate convergence.
  \item Across all methods, convergence was slower in the case of low regularization $\regparam = 1$.
\end{itemize}

\begin{figure}[tb] 
    \centering
    \subfloat[\gls{svrg}]{%
        \includegraphics[width=0.99\linewidth]{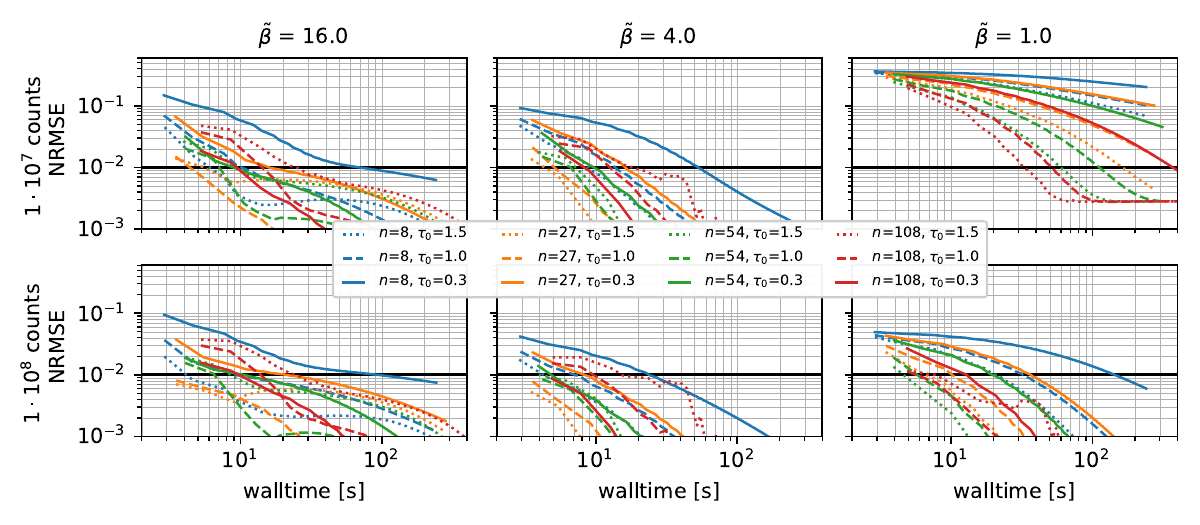}%
        \label{fig:subfig3}
    } \\
    \subfloat[\gls{saga}]{%
        \includegraphics[width=0.99\linewidth]{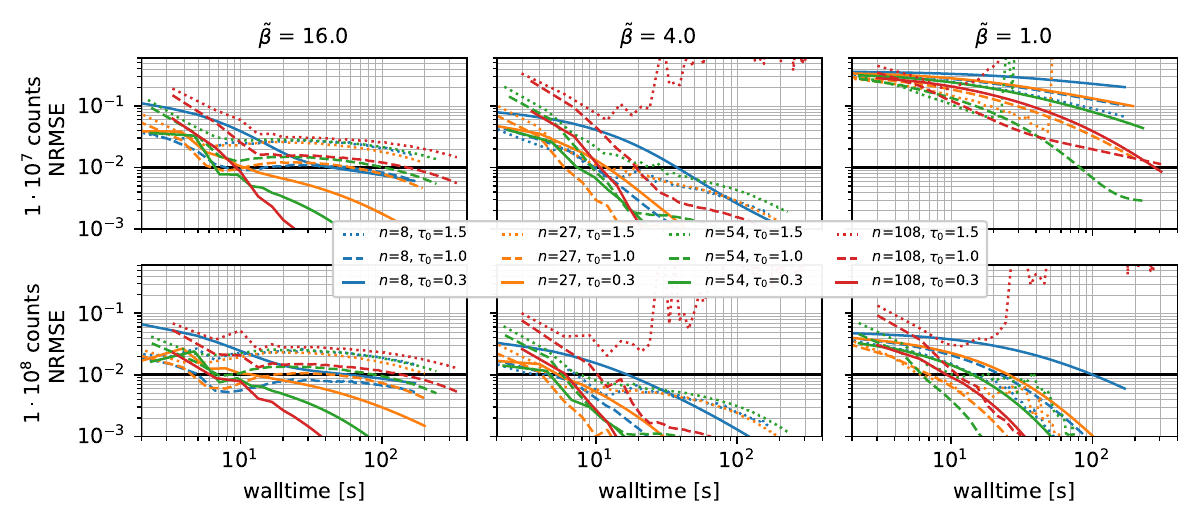}%
        \label{fig:subfig2}
    }
    \caption{Performance in terms of \gls{nrmse} versus walltime for \textbf{\gls{svrg} and \gls{saga}}, for different \textbf{number of subsets} $\nsubsets$ and \textbf{initial stepsizes} $\stepsize{0}$, using the \textbf{harmonic preconditioner}, a gentle stepsize decay with $\eta = 0.02$, 100 epochs, and subset selection without replacement.
    Results are shown for three levels of regularization $\regparam$ and two count levels. 
    For each combination of $\nsubsets$ and $\stepsize{0}$, the outcome of \textbf{1 run} is displayed.
    The thick horizontal black line shows the \gls{nrmse} target threshold of $10^{-2}$ used in the \gls{petric} challenge.}
    \label{fig:num_subsetes}
\end{figure}

\newpage

\paragraph{Impact of the number of subsets (see Fig.~\ref{fig:num_subsetes}):}
Fixing the harmonic preconditioner and vanishing stepsize rule $\stepsize{0}=1, \eta=0.02$, we varied the number of subsets $\nsubsets \in \{8,27,54,108\}$:
\begin{itemize}
  \item \gls{svrg} achieves optimal walltime convergence at $\nsubsets=27$ under medium to high $\regparam$. Lower $\regparam$ benefit from using     a greater number of subsets.
  \item Optimal values of $n$ and $\tau^{(0)}$ for \gls{saga} depend strongly on $\tilde{\beta}$: high $\tilde{\beta}$ favors a larger number of subsets with smaller $\tau^{(0)}$, medium $\tilde{\beta}$ favors $n = 27$ with $\tau^{(0)} \approx 1$, and low $\tilde{\beta}$ favors $n \approx 54$.
  \item Overall, \gls{svrg} with optimized settings achieves  faster convergence compared to \gls{saga} with optimized settings.
\end{itemize}

\begin{figure}[tb]
    \centering
    \subfloat[\gls{svrg}]{%
        \includegraphics[width=0.99\linewidth]{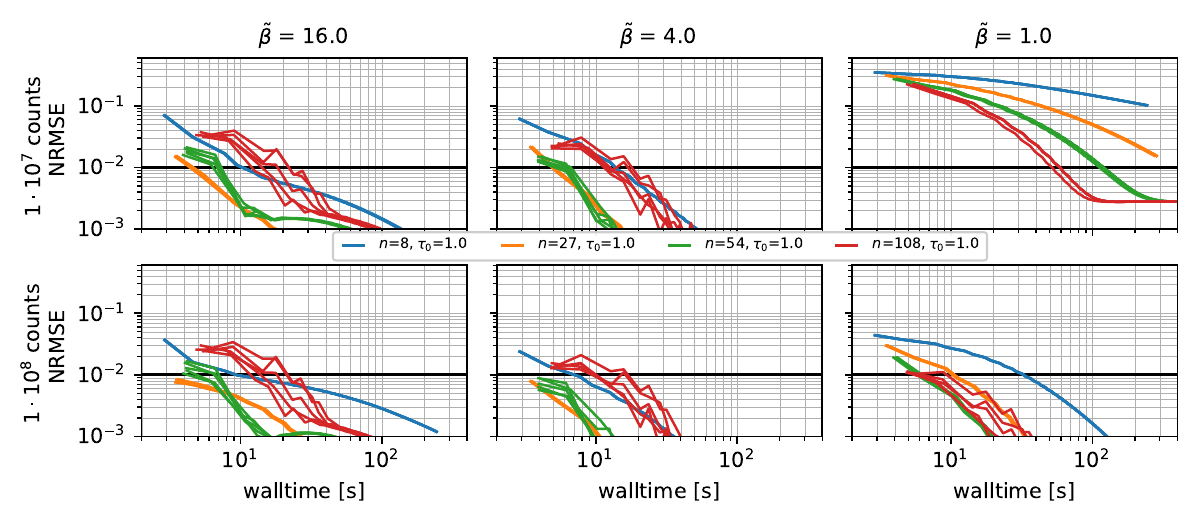}%
        \label{fig:num_subsets_runs_SVRG}
    }\\
    \subfloat[\gls{saga}]{%
        \includegraphics[width=0.99\linewidth]{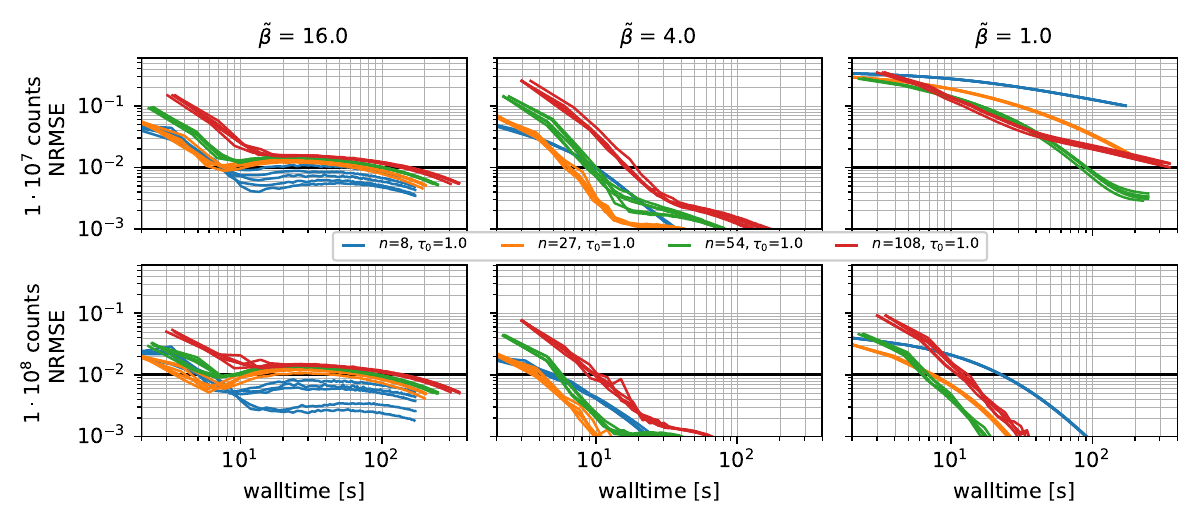}%
        \label{fig:num_subsets_runs_SAGA}
    }
    \caption{Same as Fig.~\ref{fig:num_subsetes} showing the results of \textbf{5 runs} 
    a different random seed for the subset selection.}
    \label{fig:num_subsetes_runs}
\end{figure}

\paragraph{Stability across repeated runs using different subsets orders (see Fig.~\ref{fig:num_subsetes_runs}):}
We run five independent runs (changing the random seed used for the random subset selection) of the reconstructions using \gls{svrg}, the harmonic preconditioner, $\stepsize{0} = 1$, $\eta=0.02$ and $\nsubsets \in \{8,27,54,108\}$.
The run-to-run \gls{nrmse} variation is small, especially at $\nsubsets = 27$, confirming low variance introduced by the stochastic subset selection in this setting.

\begin{figure}
    \centering
    \subfloat[non-\gls{tof} reconstructions]{%
        \includegraphics[width=0.99\linewidth]{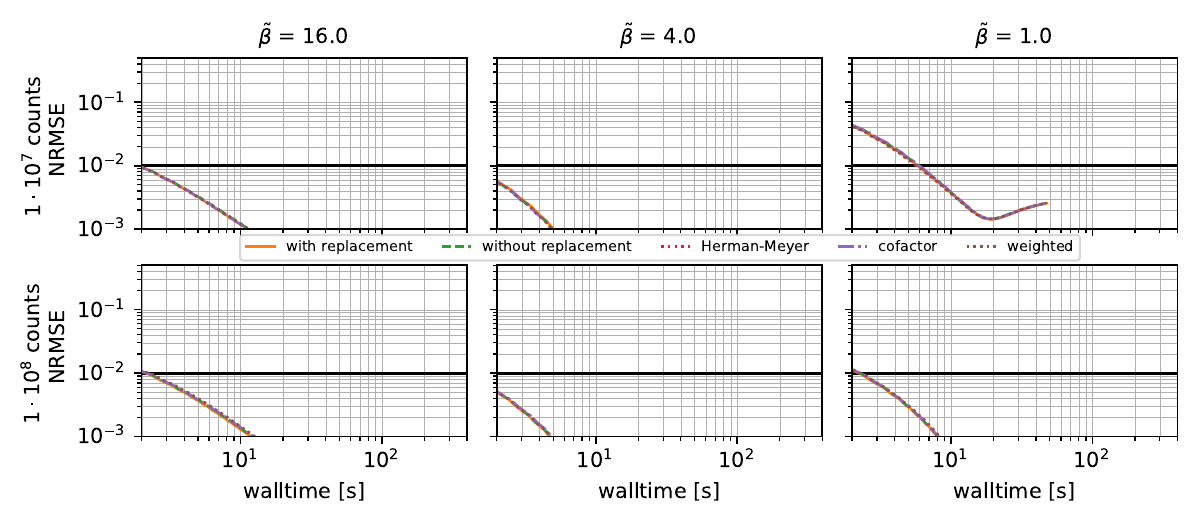}%
        \label{fig:sampling_nontof}
    }\\
    \subfloat[\gls{tof} reconstructions]{%
        \includegraphics[width=0.99\linewidth]{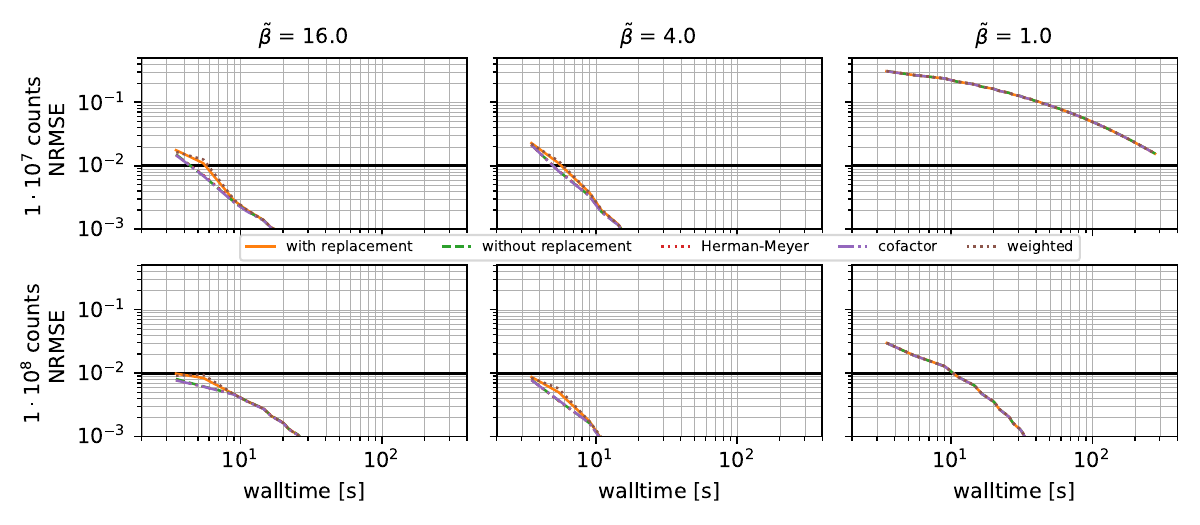}%
        \label{fig:sampling_tof}
    }\\
    \caption{Same as Fig.~\ref{fig:num_subsetes} (\gls{svrg} only) showing the results for different subset sampling strategies, $\nsubsets = 27$ subsets, the harmonic preconditioner, an initial stepsize $\stepsize{0} = 1$ and gentle stepsize decay using $\eta = 0.02$ for non-\gls{tof} (top) and \gls{tof} reconstructions (bottom).}
    \label{fig:sampling}
\end{figure}

\paragraph{Subset sampling strategy (see Fig.~\ref{fig:sampling}):}
Comparing Herman--Meyer order, uniform sampling at random with and without replacement, importance sampling, and cofactor strategies for selecting the order of subsets for \gls{svrg} with $\stepsize{0}=1$, $\nsubsets=27$, $\eta=0.02$, we observe negligible differences between all subset selection rules in simulated scenarios, with some minor benefits for sampling without replacement and cofactor sampling.

\begin{figure}
    \centering
    \subfloat[non-\gls{tof} reconstructions]{%
        \includegraphics[width=0.99\linewidth]{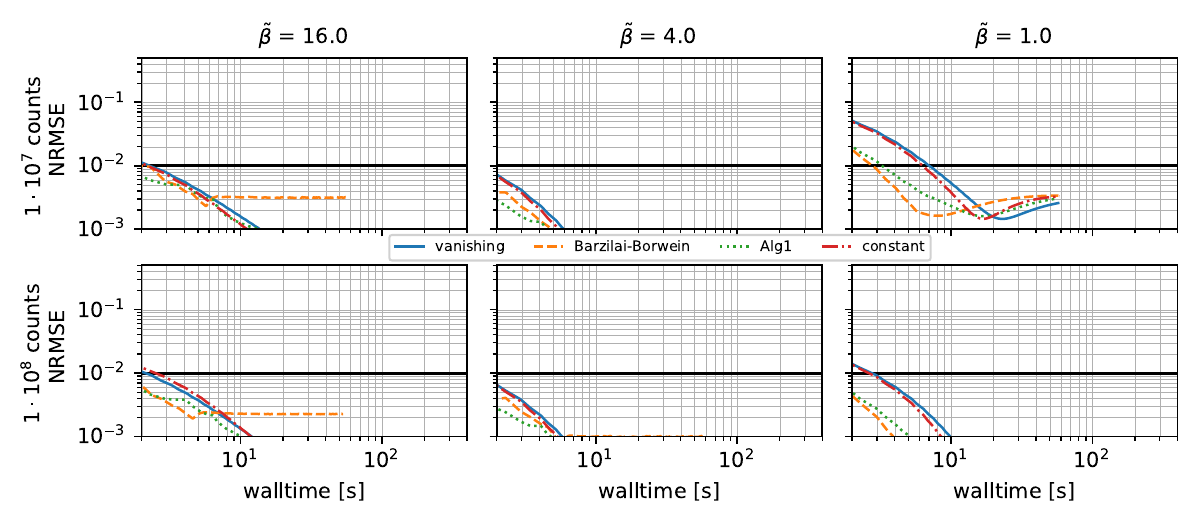}%
        \label{fig:stepsize_nontof}
    }\\
    \subfloat[\gls{tof} reconstructions]{%
        \includegraphics[width=0.99\linewidth]{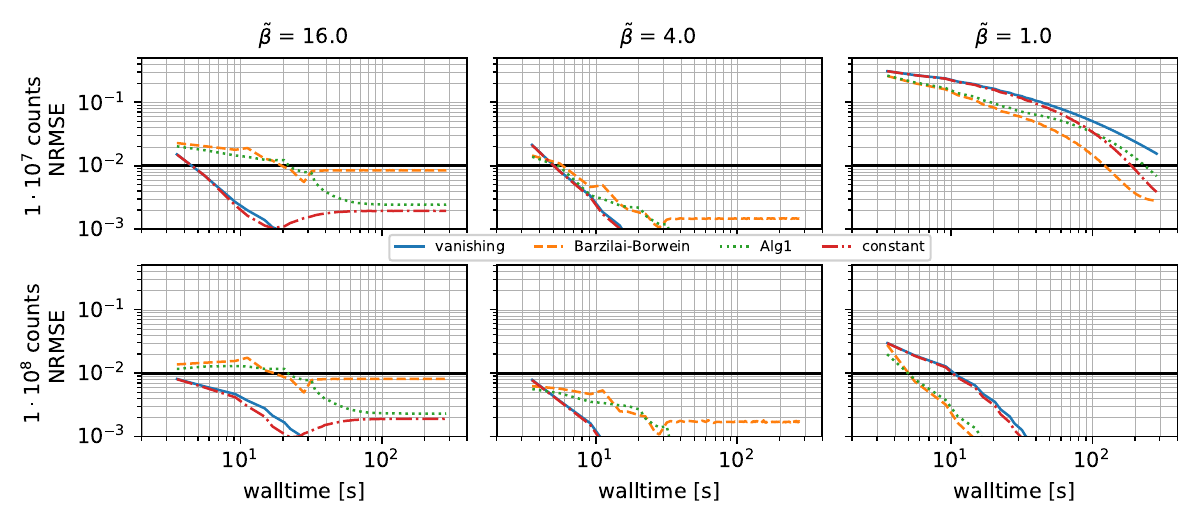}%
        \label{fig:stepsize_tof}
    }\\
    \caption{Same as Fig.~\ref{fig:sampling} showing the results for different stepsize strategies, 27 subsets, and the harmonic preconditioner for non-\gls{tof} (top) and \gls{tof} reconstructions (bottom).}
    \label{fig:stepsize}
\end{figure}

\paragraph{Stepsize rules (see Fig.~\ref{fig:stepsize}).}
We see that for \gls{svrg}, $\nsubsets=27$, and the harmonic preconditioner:
\begin{itemize}
  \item At low $\regparam$, adaptive rules (short-form \gls{bb} or heuristic ALG1) modestly outperform a simple decay.
  \item However, in the medium-to-high $\regparam$ regime, a constant or decaying initialization $\stepsize{0}=1$ yields superior \gls{tof} reconstruction performance compared to adaptive \gls{bb} schemes.
  \end{itemize}

\subsection{Simulation‐derived Conclusions} 

The inverse-crime simulation study motivated the design of our algorithms submitted to the \gls{petric} challenge in the following way:
\begin{itemize}
  \item The \textbf{harmonic-mean preconditioner} was essential to achieve stable convergence with $\stepsize{0} \approx 1$ across count and regularization regimes.
  \item \textbf{\gls{svrg}} slightly outperformed \gls{saga} in robustness and speed, and both outperformed \gls{sgd}.
  \item A moderate number of subsets, $\nsubsets \approx 27$, leads to the fastest convergence times.
\end{itemize}

These guidelines directly informed our implementation choices for the three submitted algorithms explained in detail in the next section.

%%%%%%%%%%%%%%%%%%%%%%%%%%%%%%%%%%%%%%%%%%%%%%%%%%%%%%%%%%%%%%%%%%%%%%%%%%%%%

\section{Submitted Algorithms and Their Performance}\label{sec:submitted_algorithms}

Based on the insights gained from the inverse-crime simulations in the previous section, we implemented and submitted three closely related algorithms (termed \textbf{ALG1}, \textbf{ALG2}, and \textbf{ALG3}) to the \gls{petric} challenge under the team name MaGeZ. 
All three algorithm use \gls{svrg} as the underlying stochastic gradient algorithm and apply the harmonic-mean preconditioner~\eqref{eqn:preconditioner}.
Pseudo-code that forms the basis of all three algorithms is given in Algorithm \ref{alg:svrg} in Appendix \ref{sec:app_alg}.

The available \gls{petric} training datasets were primarily used to fine-tune the algorithm hyperparameters, namely (i) number of subsets, (ii) subset selection strategy, (iii) stepsize rule and (iv) update-frequency of the preconditioner.
These are the only distinguishing features among the submitted algorithms and our choices are summarized in Table~\ref{tab:algs}.
ALG1 and ALG2 use the number of subsets as the divisor of the number of view closest to 25. ALG3 further modifies the subset count slightly using the divisor closest to 24.2 (with the goal of selecting a smaller number of subsets in some of the training datasets).
In ALG1 and ALG2 subsets are chosen uniformly at random without replacement in each iteration of each epoch. ALG3 uses the proposed cofactor rule.
ALG1 updates the preconditioner at the start of epochs 1, 2, and 3. ALG2 and ALG3 update the preconditioner at the start of epochs 1, 2, 4, and 6.
ALG1 uses a fixed, piecewise stepsize schedule, while ALG2 and ALG3 employ a short \gls{bb} rule for adaptive stepsize reduction, which is computed at the start of epochs 1, 2, 4, and 6.

\begin{table}[tb]
  \centering
  \sffamily
  \small
  \caption{Key hyperparameters of the three submitted algorithms}
  \label{tab:algs}
  \begin{tabular}{@{}p{3.3cm} p{3.5cm} p{4.3cm} p{2.5cm}@{}}
    \toprule
    & \textbf{ALG1} 
    & \textbf{ALG2} 
    & \textbf{ALG3} \\
    \midrule
    \textbf{Gradient estimator}
    & \gls{svrg}
    & same as ALG1
    & same as ALG1 \\
    \addlinespace[0.5em]
    \textbf{Preconditioner}
    & Harmonic mean
    & same as ALG1
    & same as ALG1 \\
    \addlinespace[0.5em]
    \textbf{Preconditioner update epochs}
    & 1, 2, 3
    & 1, 2, 4, 6
    & 1, 2, 4, 6 \\
    \addlinespace[0.5em]
    \textbf{Number of subsets}
    & Divisor of number of views closest to 25
    & same as ALG1
    & Divisor of number of views closest to 24.2 \\
    \addlinespace[0.5em]
    \textbf{Subset selection rule}
    & fixed random sequence without replacement
    & same as ALG1
    & cofactor \\
    \addlinespace[0.5em]
    \textbf{Stepsize rule 
    }
    & 
    $\begin{cases}
     3 & k < 10 \\
     2 & 10 \leq k < 100 \\
     1.5 & 100 \leq k < 200 \\
     1 & 200 \leq k < 300 \\
     0.5 & 300 \leq k
    \end{cases}$
    &
    $\begin{cases}
     \min(\stepsize{k}_{\rm bb},3)  &k < 10 \\
     \min(\stepsize{k}_{\rm bb},2.2) &10 \leq k < 2\nsubsets \\
     \min(\stepsize{k}_{\rm bb},1) &2\nsubsets \leq k
    \end{cases}$

    \bigskip
    
    with $\stepsize{k}_{\rm bb}$ the short \gls{bb} step
    calculated at the end of epochs 2, 4, and 6.
    & same as ALG2 \\
  \end{tabular}
\end{table}

\subsection{Performance on \gls{petric} Test Datasets}

\begin{figure}
    \centering
    \subfloat[DMI4 NEMA test dataset]{%
        \includegraphics[width=0.95\linewidth]{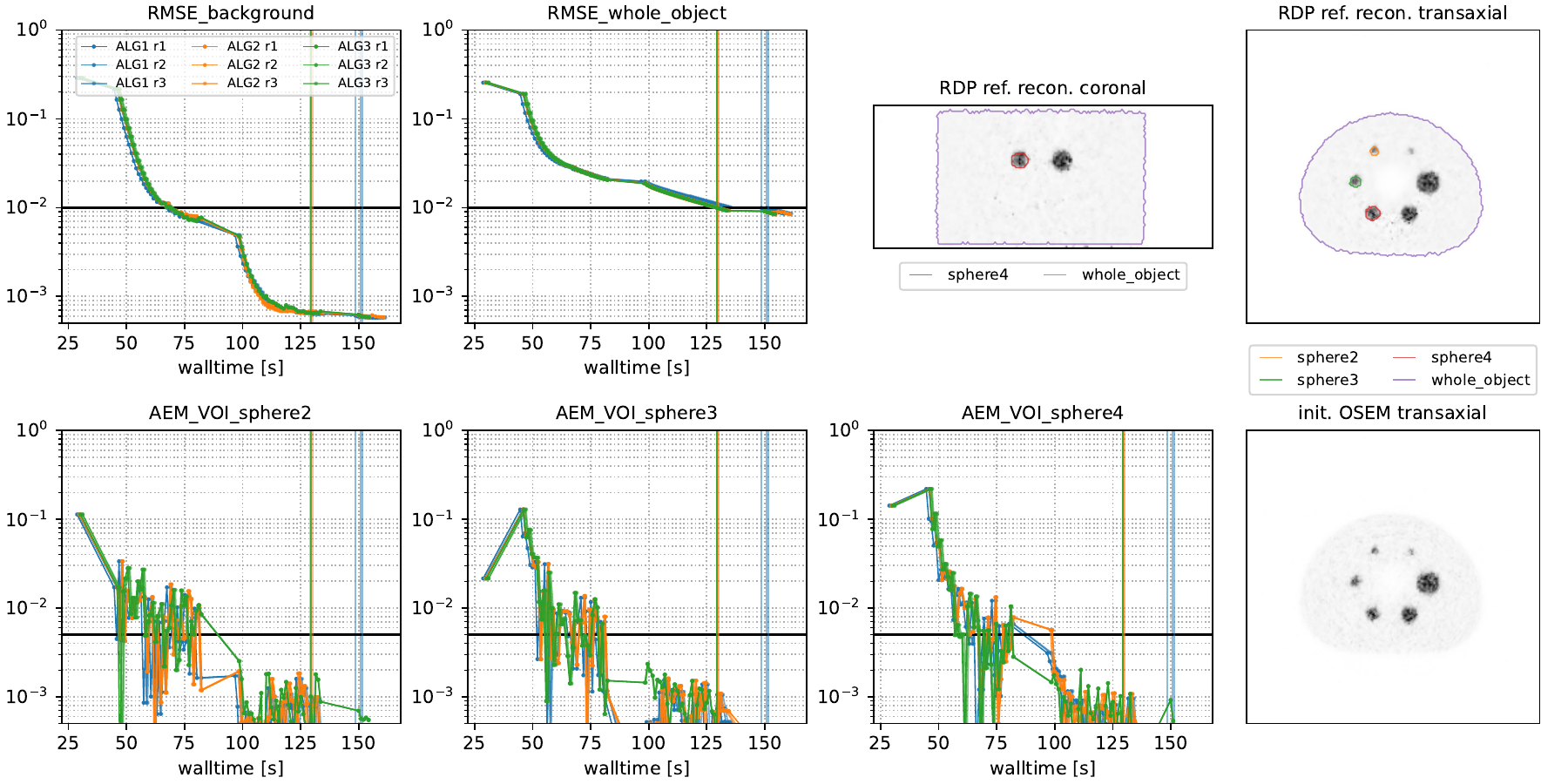}%
        \label{fig:petric_DMI4}
    }\\
    \subfloat[NeuroLF Esser test dataset]{%
        \includegraphics[width=0.95\linewidth]{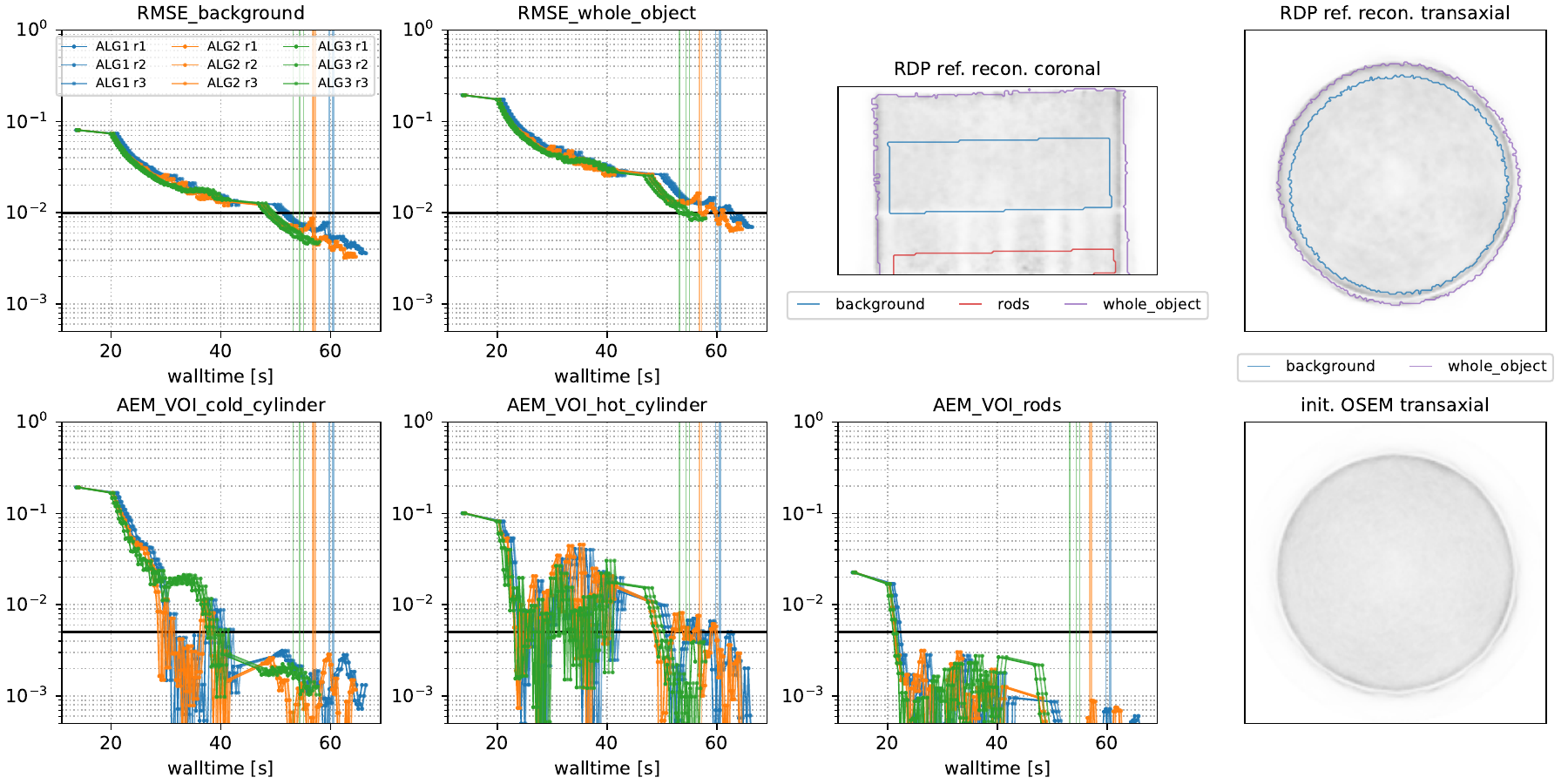}%
        \label{fig:petric_Esser}
    }\\
    \subfloat[Vision600 Hoffman test dataset]{%
        \includegraphics[width=0.95\linewidth]{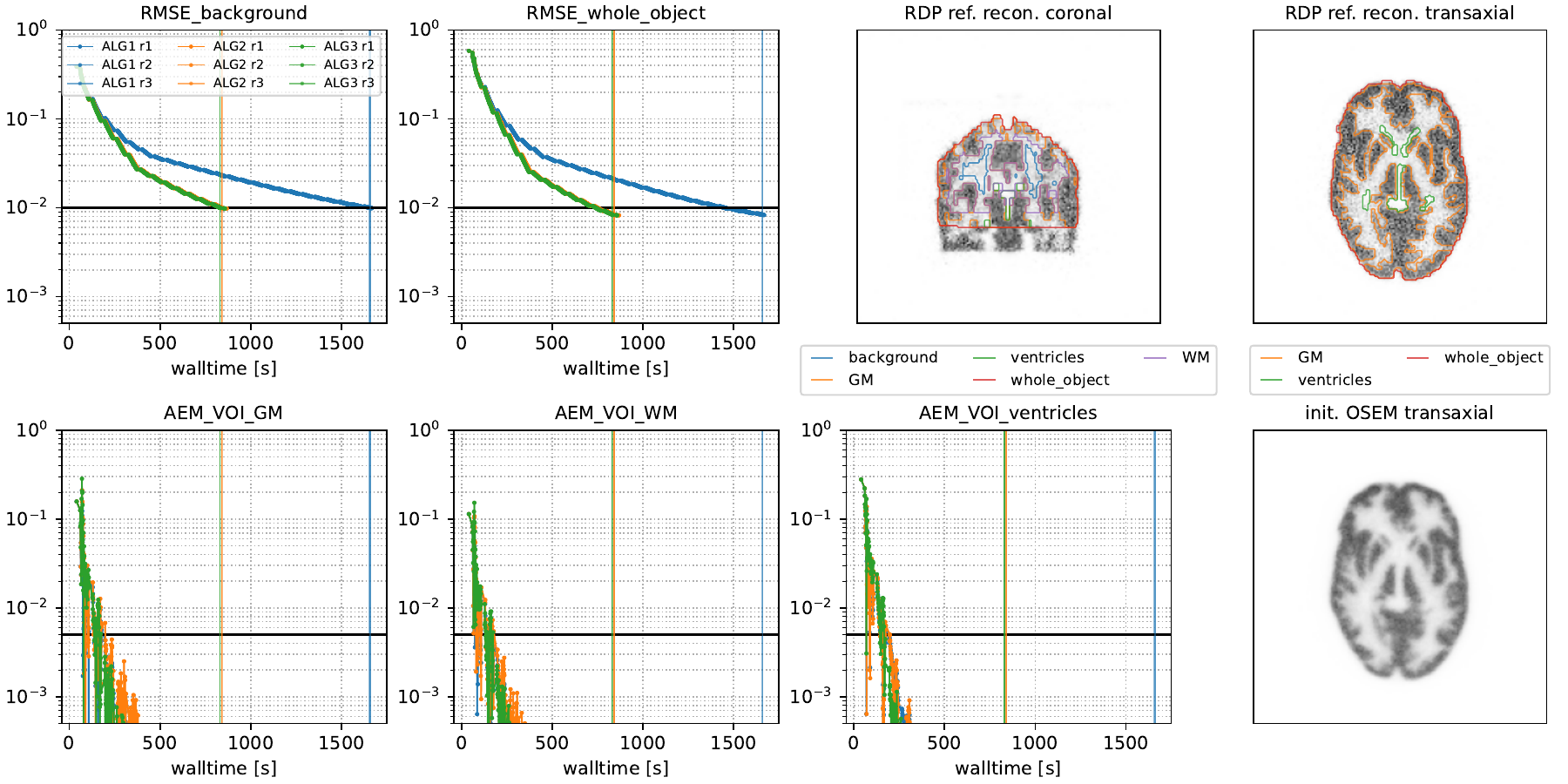}%
        \label{fig:petric_vision_hoffman}
    }\\
    \caption{Performance metrics of our three submitted algorithms evaluated on 3 representative \gls{petric} test datasets using 3 repeated runs.
    The vertical lines indicate the time when the threshold of all metrics were reached.
    Note the logarithmic scale on the y-axis and the linear scale on the x-axis.
    The top right images show a coronal and transaxial slice of the reference reconstruction alongside contour lines of the volumes of interest used for the metrics.
    The bottom right image shows the same transaxial slice of the \gls{osem} reconstruction used for initialization of all algorithms.}
    \label{fig:petric_test_1}
\end{figure}

\begin{figure}
    \centering
    \subfloat[D690 NEMA test dataset]{%
        \includegraphics[width=0.95\linewidth]{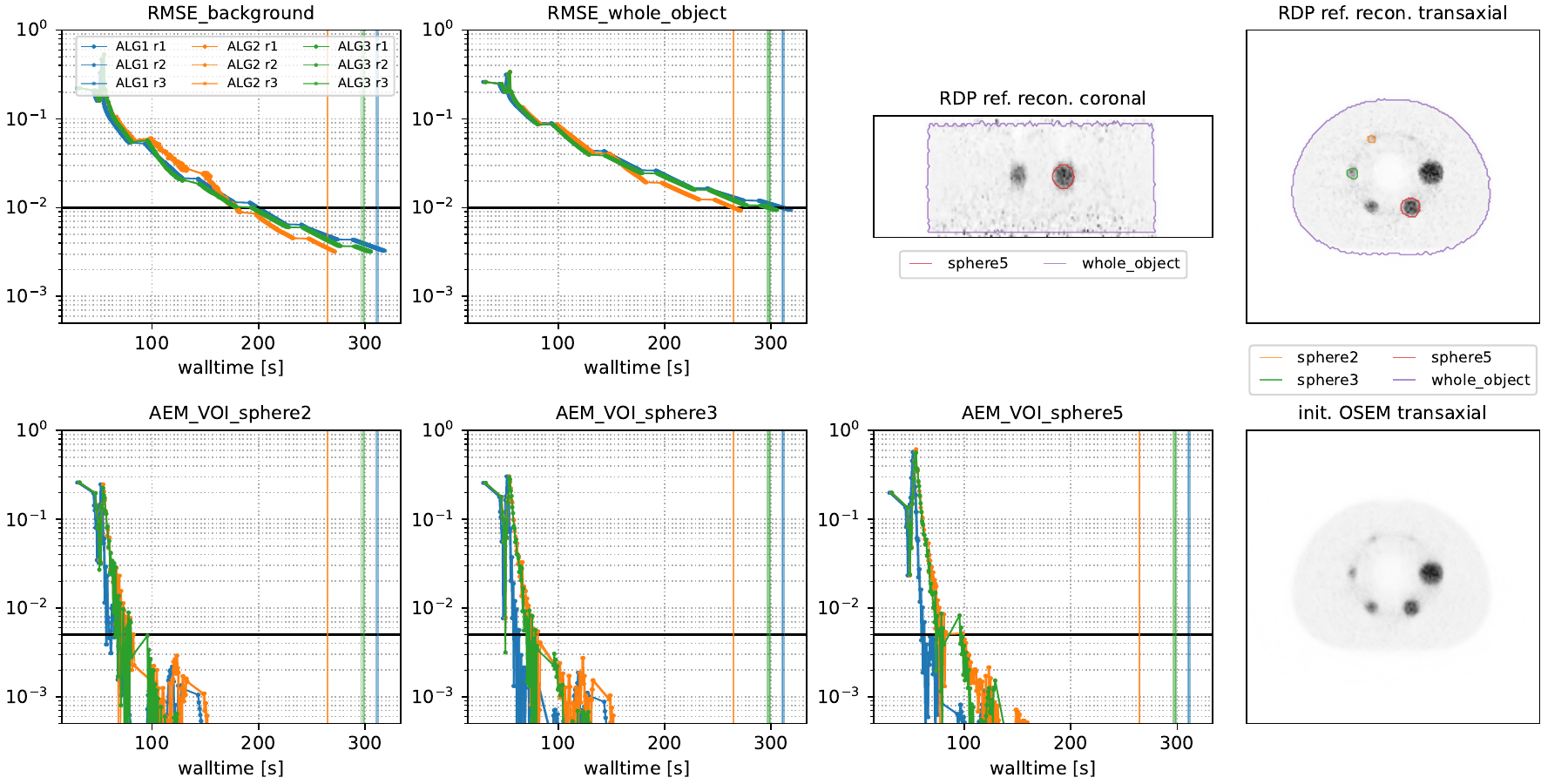}%
        \label{fig:petric_D690}
    }\\
    \subfloat[Mediso low count test dataset]{%
        \includegraphics[width=0.95\linewidth]{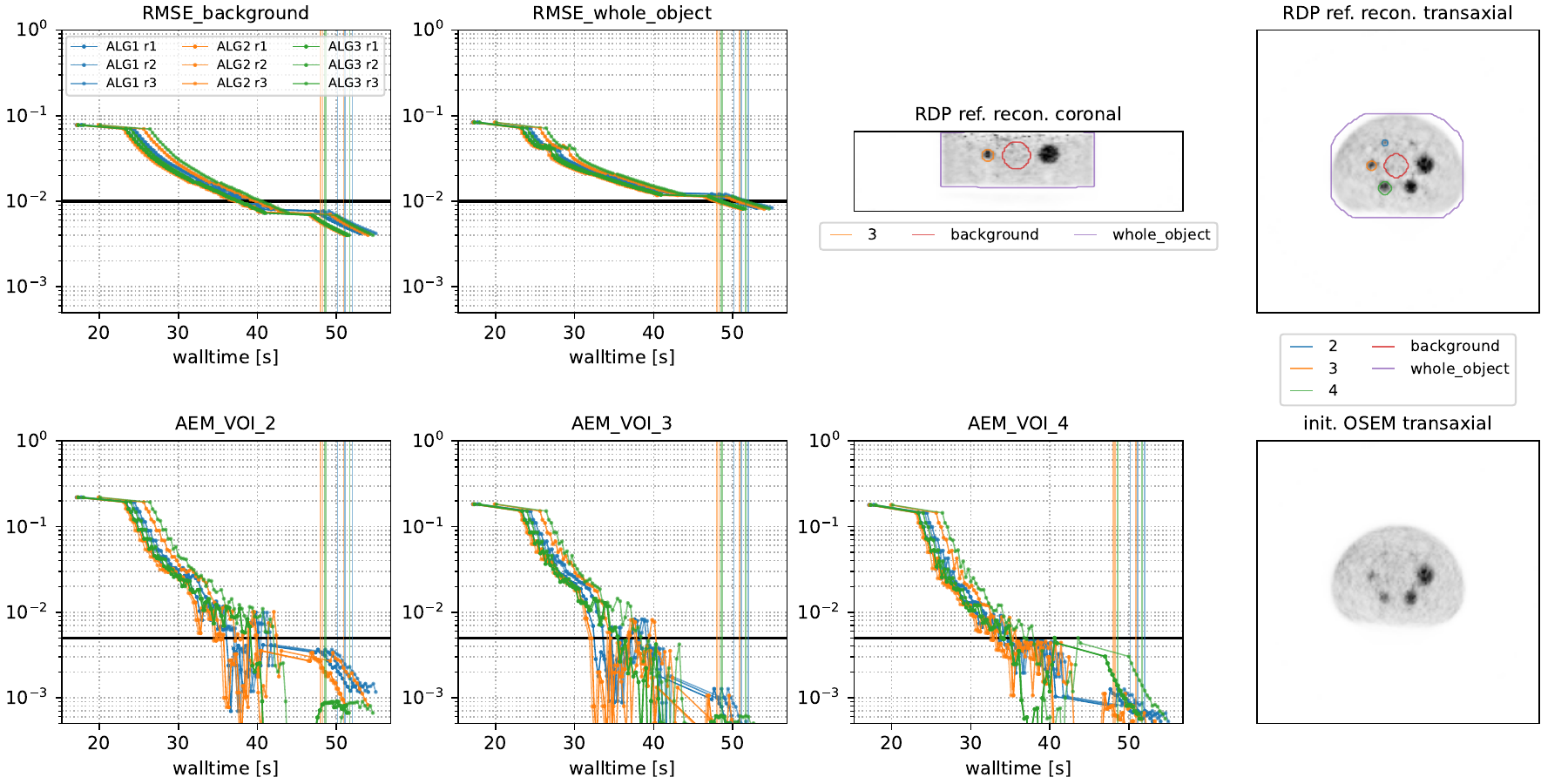}%
        \label{fig:petric_Mediso_low}
    }
    \caption{Same as Fig.~\ref{fig:petric_test_1} for 2 more datasets.}
    \label{fig:petric_test_2}
\end{figure}

Figures~\ref{fig:petric_test_1} and \ref{fig:petric_test_2} present the 
convergence behavior of all three submitted algorithms in terms of whole-object \gls{nrmse}, background \gls{nrmse}, and multiple volume-of-interest (VOI) mean absolute error metrics (AEM). 
Each dataset was reconstructed three times with all three algorithms using
a local an NVIDIA RTX A4500 GPU. 
From the two figures, we observe:
\begin{itemize}
    \item \textbf{All algorithms converge} reliably across all datasets and runs.
    \item \textbf{ALG2 and ALG3 perform similarly}, and both slightly outperform ALG1 in most cases. In the Vision600 Hoffman dataset, ALG1 almost takes twice along to reach the convergence threshold compared to ALG2 and ALG3.
    \item \textbf{For the DMI4 NEMA, NeuroLF Esser, and Mediso low-count datasets}, convergence is reached very quickly both in terms of walltime and epoch count, typically within 4 epochs.
    \item \textbf{Vision600 Hoffman dataset} shows the slowest convergence, requiring more than 23 epochs (594 updates) for ALG2 and ALG3, and more than 47 epochs (1184 updates) for ALG1.
    \item \textbf{Inter-run variability} is low; timing differences between runs are within 1–2 seconds.
    \item Across all datasets, \textbf{whole-object \gls{nrmse} is the slowest metric to converge}, becoming the bottleneck in determining the final convergence time.
\end{itemize}

Closer inspection of the stepsize behavior on the Vision600 Hoffman dataset reveals that the slower convergence of ALG1 is due to its lower final stepsize, implemented as a ``safety feature''.
After 300 updates, ALG1 reduces $\stepsize{k}$ to 0.5, whereas ALG2 and ALG3 continue to use $\stepsize{k} = 1.0$, since the \gls{bb}-based calculated adaptive stepsizes were  bigger in this dataset. 
This difference explains the kink observed in ALG1’s convergence curves around 450\,s.

\section{Discussion}\label{sec:discussion}

We now want to turn to a discussion on what we believe are important and interesting aspects of this work.

In our view, by far the most important feature of our algorithms is the improved preconditioner taking into account Hessian information of the regularizer. This meant that the stepsize choices generalized better across a range of scanners, objects, noise levels and regularization strengths.
We settled on \gls{svrg} for our gradient estimator, this choice is not as clear and may be different for other variants of the reconstruction problem. 
In our experience, employing a sophisticated method to control variance is important, but the specific approach used (e.g., \gls{svrg} or \gls{saga}) appears to be less critical. 
In contrast, other factors like stepsizes and sampling strategies seem to have a relatively minor impact, as the algorithms are not particularly sensitive to these choices.

A key aspect in our approach was to consider what can be effectively computed and what cannot. For the \gls{rdp} it is easy to compute the gradient and the diagonal Hessian, but other operations such as the proximity operator or the full Hessian are much more costly.
Similarly, the ideal number of subsets is largely a computational efficiency question. It has been observed numerous times that theoretically fewer epochs are needed with larger number of subsets. However, practically this means that the overhead per epoch increases, e.g., as the gradient is computed in each iteration of the epoch. These have the be traded off against each other.

Speaking of the \gls{rdp}, we noticed a couple of interesting features which we have not exploited in our work. First, the diagonal Hessian of the \gls{rdp} is very large in the background where the activity is small. Second, while its gradient has a Lipschitz constant, similar to the total variation and its smoothed variants, algorithms which do not rely on gradients might be beneficial. 

Between the three algorithms ALG2 and ALG3 consistently performed either similar or better than ALG1. Comparing them to the submissions of other teams it is worth noting that for almost all datasets, they perform far better than any of the other competitors which lead to MaGeZ winning the challenge overall \cite{petricleaderboard}.

Coordination between simulation insights and algorithm design was essential to our approach.
Local testing allowed us to validate generalization before submission.
Across datasets, we favored robustness over aggressive tuning.
Refinement came from iterative testing rather than theoretical guarantees alone.
Above all, our goal was to develop an algorithm that performs well out-of-the-box.

\section{Conclusions}\label{sec:conclusions}

In this paper we presented our strategy and thought process behind designing the winning strategy for the 2024 \gls{petric} challenge. We identified the key parameters for PET image reconstruction algorithms via realistic yet very fast simulations. The harmonic mean preconditioner helped to overcome the biggest roadblock of the challenge which was the tuning of parameters for a variety of settings with various scanner models, phantoms and regularization strengths.

\section*{Methods and Materials}
We do not use results from animal or human subject research. 
We use computer simulated data, with simulated scanners and measurements for the majority of the results, which can be found on \href{https://zenodo.org/records/15411476}{zenodo}.
Also, publicly available PET \href{https://github.com/SyneRBI/PETRIC}{data} provided by \gls{petric} is used.

\section*{Acknowledgments}
We acknowledge support from the EPSRC: MJE (EP/Y037286/1, EP/S026045/1,
EP/T026693/1, EP/V026259/1) and ZK (EP/X010740/1). GS acknowledges the support from NIH grant R01EB029306 and FWO project G062220N.
%%%%%%%%%%%%%%%%%%%%%%%%%%%%%%%%%%%%%%%%%%%%

\appendix

\section{Appendix}

\subsection{Gradient and Hessian of the \gls{rdp}} \label{appendix:rdp}

For completeness, we present here the first and second derivatives of the \gls{rdp} \eqref{eqn:rdp}, i.e., the gradient and the diagonal of the Hessian. Both of these are used in our proposed solution.
    
Let $d_{i,j} = \image_i - \image_j$, $s_{i,j} = \image_i + \image_j$ and $\phi_{i,j} = s_{i,j} + \gamma |d_{i,j}| + \varepsilon$. Then the first derivative is given by 
\begin{align*}
    \partial_{\image_i} \smoothness(\image) = \sum_{j \in N_i} w_{i,j} \kappa_i \kappa_j \frac{d_{i,j} (2 \phi_{i,j} - (d_{i,j} + \gamma |d_{i,j}|))}{\phi_{i,j}^2},
\end{align*}
and the second by
\begin{align*}
    \partial^2_{\image_i} \smoothness(\image) = 2 \sum_{j \in N_i} w_{i,j} \kappa_i \kappa_j \frac{(s_{i,j} - d_{i,j} + \varepsilon)^2}{\phi_{i,j}^3}.
\end{align*}

\subsection{Pseudocode for submitted preconditioned \gls{svrg} algorithm
\label{sec:app_alg}}

\begin{algorithm}
  \caption{Preconditioned \gls{svrg} Algorithm}
  \label{alg:svrg}
  \begin{algorithmic}[1]
    \Require initial image: \(\image\), number of subsets: \(\nsubsets\), 
    stepsize rule: \texttt{stepsize}, 
    sampling rule: \texttt{subset}, 
    diagonal preconditioner rule: \texttt{preconditioner},
    list of iterations to update the preconditioner: \texttt{update\_pc\_iters},
    update gradient at anchor point every $\omega$ epochs (default=2)
    \For{\(k = 0,1,\dots\)}
        \If{\(k \in \text{\texttt{update\_pc\_iters}}\)}
        \State \(\precond \gets \text{\texttt{preconditioner}}(\image)\) \Comment{update preconditioner via \eqref{eqn:preconditioner}}
      \EndIf

      \If{\(k \bmod (\omega \nsubsets) = 0\)}
        \For{\(i=1\) \textbf{to} \(\nsubsets\)} 
          \State \(\svrghist{i} \gets \nabla \objective_i(\image)\) \Comment{calculate all subset gradients at snapshot image}
        \EndFor
        \State \(\svrghist{} \gets \sum_{i=1}^\nsubsets \svrghist{i}\)
        \State \(\tilde \nabla \gets \svrghist{}\)
      \Else
        \State \(i \gets \text{\texttt{subset}} (k)\) 
        \State \(\tilde \nabla \gets \nsubsets \bigl(\nabla \objective_i(\image) - \svrghist{i}\bigr) + \svrghist{}\)
      \EndIf
        \State \(\tau \gets \text{\texttt{stepsize}} (k)\) 
      \State \(\image \gets \image -\tau\precond\tilde \nabla\)
      \If{stopping criterion is reached} 
        \Return{$\image$}
       \EndIf
       \EndFor
  \end{algorithmic}
\end{algorithm}

%%%%%%%%%%%%%%%%%%%%%%%%%%%%%%%%%%%%%%%%%%%%%%%%%%%%%%%%%%%%%%%%%%%%%%%%%%%%%%%%%%%%%%%

\clearpage

\printnoidxglossaries

%--------------------------------------------------------------------

\bibliographystyle{unsrt}    % or another style

\begin{thebibliography}{10}

\bibitem{nuyts2002concave}
Johan Nuyts, Dirk Bequ{\'e}, Patrick Dupont, and Luc Mortelmans.
\newblock A concave prior penalizing relative differences for maximum-a-posteriori reconstruction in emission tomography.
\newblock {\em IEEE Transactions on nuclear science}, 49(1):56--60, 2002.

\bibitem{petric}
Casper da~Costa-Luis, Matthias~J. Ehrhardt, Christoph Kolbitsch, Evgueni Ovtchinnikov, Edoardo Pasca, Kris Thielemans, and Charalampos Tsoumpas.
\newblock Petric: Pet rapid image reconstruction challenge, 2025.

\bibitem{teoh2015}
Eugene~J. Teoh, Daniel~R. McGowan, Ruth~E. Macpherson, Kevin~M. Bradley, and Fergus~V. Gleeson.
\newblock Phantom and clinical evaluation of the bayesian penalized likelihood reconstruction algorithm q.clear on an lyso pet/ct system.
\newblock {\em Journal of Nuclear Medicine}, 56(9):1447--1452, July 2015.

\bibitem{teoh2015b}
Eugene~J. Teoh, Daniel~R. McGowan, Kevin~M. Bradley, Elizabeth Belcher, Edward Black, and Fergus~V. Gleeson.
\newblock Novel penalised likelihood reconstruction of pet in the assessment of histologically verified small pulmonary nodules.
\newblock {\em European Radiology}, 26(2):576--584, May 2015.

\bibitem{Ahn2015}
Sangtae Ahn, Steven~G Ross, Evren Asma, Jun Miao, Xiao Jin, Lishui Cheng, Scott~D Wollenweber, and Ravindra~M Manjeshwar.
\newblock Quantitative comparison of osem and penalized likelihood image reconstruction using relative difference penalties for clinical pet.
\newblock {\em Physics in Medicine and Biology}, 60(15):5733--5751, July 2015.

\bibitem{ahn2003globally}
Sangtae Ahn and Jeffrey~A Fessler.
\newblock Globally convergent image reconstruction for emission tomography using relaxed ordered subsets algorithms.
\newblock {\em IEEE transactions on medical imaging}, 22(5):613--626, 2003.

\bibitem{twyman2022investigation}
Robert Twyman, Simon Arridge, Zeljko Kereta, Bangti Jin, Ludovica Brusaferri, Sangtae Ahn, Charles~W Stearns, Brian~F Hutton, Irene~A Burger, Fotis Kotasidis, et~al.
\newblock An investigation of stochastic variance reduction algorithms for relative difference penalized 3d pet image reconstruction.
\newblock {\em IEEE Transactions on Medical Imaging}, 42(1):29--41, 2022.

\bibitem{ovtchinnikov2020sirf}
Evgueni Ovtchinnikov, Richard Brown, Christoph Kolbitsch, Edoardo Pasca, Casper da~Costa-Luis, Ashley~G Gillman, Benjamin~A Thomas, Nikos Efthimiou, Johannes Mayer, Palak Wadhwa, et~al.
\newblock Sirf: synergistic image reconstruction framework.
\newblock {\em Computer Physics Communications}, 249:107087, 2020.

\bibitem{hudson1994accelerated}
H.~Malcolm Hudson and Richard~S. Larkin.
\newblock Accelerated image reconstruction using ordered subsets of projection data.
\newblock {\em IEEE transactions on medical imaging}, 13(4):601--609, 1994.

\bibitem{ehrhardt2019faster}
Matthias~J Ehrhardt, Pawel Markiewicz, and Carola-Bibiane Sch{\"o}nlieb.
\newblock Faster pet reconstruction with non-smooth priors by randomization and preconditioning.
\newblock {\em Physics in Medicine \& Biology}, 64(22):225019, 2019.

\bibitem{kereta2021stochastic}
{\v{Z}}eljko Kereta, Robert Twyman, Simon Arridge, Kris Thielemans, and Bangti Jin.
\newblock Stochastic em methods with variance reduction for penalised pet reconstructions.
\newblock {\em Inverse Problems}, 37(11):115006, 2021.

\bibitem{schramm2022fast}
Georg Schramm and Martin Holler.
\newblock Fast and memory-efficient reconstruction of sparse poisson data in listmode with non-smooth priors with application to time-of-flight pet.
\newblock {\em Physics in Medicine \& Biology}, 67(15):155020, 2022.

\bibitem{ehrhardt2024guide}
Matthias~Joachim Ehrhardt, Zeljko Kereta, Jingwei Liang, and Junqi Tang.
\newblock A guide to stochastic optimisation for large-scale inverse problems.
\newblock {\em Inverse Problems}, 2024.

\bibitem{papoutsellis2024stochastic}
Evangelos Papoutsellis, Casper da~Costa-Luis, Daniel Deidda, Claire Delplancke, Margaret Duff, Gemma Fardell, Ashley Gillman, Jakob~S J{\o}rgensen, Zeljko Kereta, Evgueni Ovtchinnikov, et~al.
\newblock Stochastic optimisation framework using the core imaging library and synergistic image reconstruction framework for pet reconstruction.
\newblock {\em arXiv preprint arXiv:2406.15159}, 2024.

\bibitem{bredies2010total}
Kristian Bredies, Karl Kunisch, and Thomas Pock.
\newblock Total generalized variation.
\newblock {\em SIAM Journal on Imaging Sciences}, 3(3):492--526, 2010.

\bibitem{knoll2016joint}
Florian Knoll, Martin Holler, Thomas Koesters, Ricardo Otazo, Kristian Bredies, and Daniel~K Sodickson.
\newblock Joint mr-pet reconstruction using a multi-channel image regularizer.
\newblock {\em IEEE transactions on medical imaging}, 36(1):1--16, 2016.

\bibitem{ehrhardt2016pet}
Matthias~J Ehrhardt, Pawel Markiewicz, Maria Liljeroth, Anna Barnes, Ville Kolehmainen, John~S Duncan, Luis Pizarro, David Atkinson, Brian~F Hutton, Sebastien Ourselin, et~al.
\newblock Pet reconstruction with an anatomical mri prior using parallel level sets.
\newblock {\em IEEE transactions on medical imaging}, 35(9):2189--2199, 2016.

\bibitem{xie2020generative}
Zhaoheng Xie, Reheman Baikejiang, Tiantian Li, Xuezhu Zhang, Kuang Gong, Mengxi Zhang, Wenyuan Qi, Evren Asma, and Jinyi Qi.
\newblock Generative adversarial network based regularized image reconstruction for pet.
\newblock {\em Physics in Medicine \& Biology}, 65(12):125016, 2020.

\bibitem{singh2024scorebasedpet}
Imraj~RD Singh, Alexander Denker, Riccardo Barbano, {\v Z}eljko Kereta, Bangti Jin, Kris Thielemans, Peter Maass, and Simon Arridge.
\newblock Score-based generative models for pet image reconstruction.
\newblock {\em Machine Learning for Biomedical Imaging}, 2:547--585, 2024.

\bibitem{tsai2019benefits}
Yu-Jung Tsai, Georg Schramm, Sangtae Ahn, Alexandre Bousse, Simon Arridge, Johan Nuyts, Brian~F Hutton, Charles~W Stearns, and Kris Thielemans.
\newblock Benefits of using a spatially-variant penalty strength with anatomical priors in pet reconstruction.
\newblock {\em IEEE Transactions on Medical Imaging}, 39(1):11--22, 2019.

\bibitem{Defazio2014}
Aaron Defazio, Francis Bach, and Simon Lacoste-Julien.
\newblock Saga: A fast incremental gradient method with support for non-strongly convex composite objectives.
\newblock In {\em Advances in Neural Information Processing Systems}, volume~2, pages 1646--1654, 2014.

\bibitem{johnson2013svrg}
Rie Johnson and Tong Zhang.
\newblock Accelerating stochastic gradient descent using predictive variance reduction.
\newblock In C~J~C Burges, L~Bottou, M~Welling, Z~Ghahramani, and K~Q Weinberger, editors, {\em Advances in Neural Information Processing Systems}, volume~26, 2013.

\bibitem{tan2016barzilai}
Conghui Tan, Shiqian Ma, Yu-Hong Dai, and Yuqiu Qian.
\newblock Barzilai-borwein step size for stochastic gradient descent.
\newblock {\em Advances in neural information processing systems}, 29, 2016.

\bibitem{ivgi2023dog}
Maor Ivgi, Oliver Hinder, and Yair Carmon.
\newblock Dog is sgd's best friend: A parameter-free dynamic step size schedule.
\newblock In {\em International Conference on Machine Learning}, pages 14465--14499. PMLR, 2023.

\bibitem{vaswani2019painless}
Sharan Vaswani, Aaron Mishkin, Issam Laradji, Mark Schmidt, Gauthier Gidel, and Simon Lacoste-Julien.
\newblock Painless stochastic gradient: Interpolation, line-search, and convergence rates.
\newblock {\em Advances in neural information processing systems}, 32, 2019.

\bibitem{herman1993hmorder}
Gabor~T. Herman and Lorraine~B. Meyer.
\newblock Algebraic reconstruction techniques can be made computationally efficient (positron emission tomography application).
\newblock {\em IEEE Transactions on Medical Imaging}, 12(3):600--609, 1993.

\bibitem{schramm2023}
Georg Schramm and Kris Thielemans.
\newblock {PARALLELPROJ--an open-source framework for fast calculation of projections in tomography}.
\newblock {\em Frontiers in Nuclear Medicine}, Volume 3 - 2023, 2024.

\bibitem{petricleaderboard}
Casper da~Costa-Luis, Matthias~J. Ehrhardt, Christoph Kolbitsch, Evgueni Ovtchinnikov, Edoardo Pasca, Kris Thielemans, and Charalampos Tsoumpas.
\newblock Petric: Pet rapid image reconstruction challenge - leaderboard, 2025.

\end{thebibliography}

\end{document}